\documentclass[a4paper,UKenglish,cleveref,autoref,thm-restate]{lipics-v2021}

\usepackage{graphicx}

\usepackage{array}
\usepackage{paralist}

\usepackage{caption}
\usepackage{subcaption}
\usepackage{thmtools} 
\usepackage{amsmath,amssymb}
\usepackage{csquotes}
\usepackage{enumitem}
\usepackage{booktabs}
\usepackage{xspace}
\usepackage{hyperref}
\usepackage{color}
\usepackage{makecell}

\graphicspath{{figures/}}

\newcommand{\questionBox}[1]{\smallskip
    \begin{nolinenumbers}
        \noindent\fbox{\begin{minipage}{0.98\linewidth}
                \smallskip

                #1

                \smallskip
        \end{minipage}}
        \smallskip
    \end{nolinenumbers}
    
}

\newcommand{\rome}{\texttt{rome}\xspace}
\newcommand{\BA}{\texttt{BA}\xspace}

\hideLIPIcs

\title{Using Reinforcement Learning to Optimize the Global and Local Crossing Number}

\author{Timo Brand}{Technical University of Munich, Heilbronn, Germany}{timo.brand@tum.de}{https://orcid.org/0009-0004-3111-2045}{}

\author{Henry F\"orster}{John Cabot University, Rome, Italy}{henry.foerster@johncabot.edu}{https://orcid.org/0000-0002-1441-4189}{}

\author{Stephen Kobourov}{Technical University of Munich, Heilbronn, Germany}{stephen.kobourov@tum.de}{https://orcid.org/0000-0002-0477-2724}{}

\author{Daniel Kohrt}{Technical University of Munich, Garching, Germany}{daniel.kohrt@tum.de}{}{}

\author{Robin Schukrafft}{Technical University of Munich, Heilbronn, Germany}{robin.schukrafft@tum.de}{}{}

\author{Markus Wallinger}{Technical University of Munich, Heilbronn, Germany}
{markus.wallinger@tum.de}{https://orcid.org/0000-0002-2191-4413}{}

\author{Johannes Zink}{Technical University of Munich, Heilbronn, Germany}{johannes.zink@tum.de}{https://orcid.org/0000-0002-7398-718X}{}

\authorrunning{T. Brand, H. F\"orster, S. Kobourov, D. Kohrt, R. Schukrafft, M. Wallinger, J. Zink}

\Copyright{T. Brand, H. F\"orster, S. Kobourov, D. Kohrt, R. Schukrafft, M. Wallinger, J. Zink}


\nolinenumbers


\begin{document}

\maketitle

\begin{abstract}
Graph drawing concerns the algorithmic visualization of graphs.
A \emph{good} drawing of a graph is easy to read
and facilitates solving tasks on the graph.
Several properties have been identified
to occur in good drawings of graphs.
Such properties include a low number of crossings,
large angles between edges, short edges, and depicting symmetries.
Many of these properties are explicitly measurable metrics.
This lets us model a graph-drawing problem as a game
where a single player iteratively moves vertices of a straight-line graph drawing to reduce edge crossings. 
We
investigate whether reinforcement learning can discover effective strategies for playing
this game.
Our reinforcement-learning agent observes the local geometric and structural context of a vertex and selects a movement direction with the goal of reducing either the global or the local crossing number, that is,
either the total number of crossings or the maximum number of crossings per edge.
We compare the resulting strategies to existing methods and established crossing-minimization heuristics on standard benchmark graphs. 
While our approach does not out-compete state-of-the-art methods for minimizing the global crossing number,
it is competitive and often superior for minimizing the local crossing number.
\end{abstract}

\section{Introduction}
Reinforcement-learning breakthroughs in game-playing made headlines in popular media.
After AlphaGo~\cite{DBLP:journals/nature/SilverHMGSDSAPL16} beat the best human player at Go in 2016, the more general programs AlphaGoZero~\cite{DBLP:journals/nature/SilverSSAHGHBLB17} and AlphaZero~\cite{doi:10.1126/science.aar6404} mastered Go, chess, and Shogi after only being taught the rules and objectives, finding new strategies by playing against themselves.

In \emph{reinforcement learning} (RL), an agent acts in an environment based on an \emph{observation vector} that describes the environment.
Based on this vector, the agent chooses an action from the available actions, and the resulting change in the environment is measured.
Based on the measurement, the agent receives a \emph{reward}, a positive number for a ``good'' change or a negative number for a ``bad'' change, and tries to maximize its cumulative reward.
The agent explores its options by trying different actions for different observation vectors.
Internally, a neural network is typically used whose input is the observation vector and whose output is a recommendation for the next action.

RL is applicable to a variety of games,
e.g., puzzles, card and board games,
and even complex computer games~\cite{DBLP:journals/nature/MnihKSRVBGRFOPB15,DBLP:journals/corr/abs-1912-06680,DBLP:journals/nature/SchrittwieserAH20}.
We investigate the following question
that arises naturally when applying the RL game-playing paradigm to graph drawing:

\questionBox{\textit{Can we use reinforcement learning to optimize graph drawings
        with respect to a given graph drawing aesthetics measure?}}

From this perspective, graph drawing can be viewed as a single-player game
with full information, which is also called a \emph{puzzle}:
starting from an initial drawing, a player repeatedly selects a vertex and moves it in the plane, observing how this move affects the drawing quality.
This can be done on an integer grid or without restrictions to a grid.
In principle, any graph drawing metric is applicable.
Here, we use an integer grid
and we employ the maybe most natural objective
to reach a drawing with few crossings. 
This interpretation closely mirrors how humans and algorithms approached the automatic Graph Drawing Challenge,
motivating us to study graph drawing as a single-player game for an RL agent.

\subparagraph*{Our Contributions.}
We present an RL framework for graph drawing to optimize local and global crossings by post-processing a given layout; see \cref{sec:approach}.
We perform a quantitative analysis on
two graph data sets (Rome and Barabási-Albert graphs)
comparing our two models against other state-of-the-art approaches;
see \cref{sec:experiments}.
We identify strengths and weaknesses of the current approach,
which can be seen as a guide for future work on RL in graph drawing
using the game-playing paradigm;
see \cref{sec:discussion,sec:conclusion}.
All of our code, benchmark data, trained models, and results are available in a git repository.\footnote{%
\url{https://github.com/j-zink-wuerzburg/rl-4-gd-crossing-reduction}}

\section{Preliminaries and Related Work}
\label{sec:preliminaries}

A drawing of a graph assigns each vertex to a point and each
edge to a curve
whose endpoints are the points of its two vertices.
In a straight-line drawing, edges are line segments.

\subparagraph*{Crossing Number.}
For a given drawing $\Gamma$ of a graph~$G$,
the \emph{crossing number}~$\mathrm{cr}(\Gamma)$ denotes the number of pairwise edge crossings in~$\Gamma$.
The \emph{crossing number}~$\mathrm{cr}(G)$ of a graph $G$ is, among every drawing $\Gamma$ of~$G$, the minimum of $\mathrm{cr}(\Gamma)$.
An often studied variation is the \emph{rectilinear crossing number} $\overline{\mathrm{cr}}(G)$, which denotes, for a graph $G$,
the same value restricted to straight-line drawings of~$G$.
The \emph{local crossing number} $\mathrm{lcr}(\Gamma)$ of a drawing $\Gamma$ is the maximum number of edges that cross a single edge in~$\Gamma$.
The \emph{local crossing number}~$\mathrm{lcr}(G)$ and the \emph{rectilinear local crossing number}~$\overline{\mathrm{lcr}}(G)$
of a graph~$G$ are defined analogously to $\mathrm{cr}(G)$ and $\overline{\mathrm{cr}}(G)$, respectively.
Computing $\mathrm{cr}(G)$~\cite{doi:10.1137/0604033} and $\mathrm{lcr}(G)$~\cite{DBLP:journals/algorithmica/GrigorievB07,DBLP:conf/gd/KorzhikM08,DBLP:journals/jgt/KorzhikM13}
are known to be NP-complete, while $\overline{\mathrm{cr}}(G)$~\cite{bienstock,DBLP:journals/jgaa/Schaefer21a} and $\overline{\mathrm{lcr}}(G)$~\cite{DBLP:journals/jgaa/Schaefer21} are even $\exists\mathbb{R}$-complete; see also Schaefer's~\cite{survey-crossing-number} survey.
Several studies evidence that the number of edge crossings
is crucial for the readability of a graph drawing~\cite{HuangEHL13,Purchase2000,Purchase2002,Ware2002}. 
The local crossing number is a natural variation.
Moreover,
there are
many
theoretical results on drawings with $\mathrm{lcr}(\Gamma)=k$, called \emph{$k$-planar} drawings; see~\cite{DBLP:journals/csur/DidimoLM19,DBLP:journals/csr/KobourovLM17} for surveys on the topic.

In the rest of the paper, we only consider straight-line drawings
and, hence, (versions of) the rectilinear crossing number.
For brevity, we drop the word \emph{rectilinear} from now on.

\subparagraph*{Crossing Minimization Heuristics.}
Stress-based (e.g., \textsf{Kamada--Kawai}~\cite{kamada1989algorithm}) and force-directed (e.g.,
\textsf{Fruchterman--Reingold}~\cite{fruchterman1991graph}) algorithms
are well known to produce drawings with acceptable crossing numbers despite not explicitly optimizing crossings.
It has even been shown that, in random geometric graphs,
the correlation between the expected crossing number
and stress is positive~\cite{DBLP:conf/gd/ChimaniDR18}.
There are extensions of force-directed algorithms interfering with crossings~\cite{DBLP:conf/gd/DidimoLR10}
and extra forces improving angles formed at crossings and vertices~\cite{DBLP:journals/cj/ArgyriouBS13,HuangEHL13}.

Several crossing-minimization heuristics were developed
for
the annual Graph Drawing Contest,\footnote{See \url{https://mozart.diei.unipg.it/gdcontest/}.} which in 2017--2020 focused on crossings.
In 2017/18, the goal was to maximize the minimum crossing angle, addressed by simulated annealing~\cite{DBLP:conf/gd/DemelDMRW18} and by probabilistic hill climbing~\cite{DBLP:journals/cj/BekosFGHKSS21}; both iteratively pick a vertex with a small crossing angle and perturb its position.
In 2019/20, the focus shifted to crossings in \emph{upward drawings}, for which a more efficient version of the hill-climbing approach, called \textsf{Tübingen-Bus}, was proposed~\cite{maxi}.
Radermacher, Reichard, Rutter, and Wagner~\cite{radermacher2019geometric} subsequently investigated variants of the simulated annealing for crossing minimization.
They provide several \emph{local optimization}-based geometric heuristics, among others \textsf{Vertex Movement} and \textsf{Edge Insertion}, where the latter has also been proposed before~\cite{DBLP:journals/jss/BatiniTT84}.
\textsf{Vertex Movement} moves each vertex to its crossing-minimal position in the current drawing, while \textsf{Edge Insertion} builds the drawing edge by edge starting from a planar subgraph, using \textsf{Vertex Movement} as a subroutine to keep added crossings low.

Also researchers from the \textsf{University of Arizona}
competed in the GD contests
and published their optimization algorithm combining linear programming and gradient descent~\cite{DBLP:conf/gd/DevkotaALIK19}
to simultaneously optimize stress and the number of crossings.
In a later version called
\textsf{(SGD)\textsuperscript{2}}~\cite{ahmed2022multicriteria},
the gradient restricted to a subset of the vertices is computed for a set of objective functions, and
the layout is updated by moving vertices in the direction maximizing the gradient.

Recently, there has been some interest in using neural networks to compute
drawings~\cite{DBLP:conf/gd/GiovannangeliLA21,DBLP:journals/tvcg/KwonM20,DBLP:journals/cga/WangYHS21,DBLP:journals/tvcg/WangJWCMQ20}.
Because previous approaches struggled to optimize non-differentiable objective functions -- like the number of crossings -- Wang, Yen, Hu, and Shen~\cite{wang2023smartgd} introduced \textsf{SmartGD}, a Generative Adversarial Network (GAN)-based deep learning framework.

\subparagraph*{Reinforcement Learning.}

There are several
frameworks designed for playing arbitrary games, e.g., Google DeepMind's \textit{OpenSpiel}~\cite{LanctotEtAl2019OpenSpiel} and \textit{MuZero}~\cite{DBLP:journals/nature/SchrittwieserAH20,muzero-general},
\textit{Ai Ai}~\cite{Tavener2025}, \textit{AmzPlayer}~\cite{amzplayer2023}.
In these frameworks, the rules of the games,
the observation vector, the possible actions, and the reward function
must be specified.
A sensible encoding of the environment into
observation vectors, possible actions, and rewards
is crucial for sensible outcomes in a reasonable amount of time; e.g., when encoding graphs, a learning algorithm
should not base its decisions on the order of the labels
of the graph vertices but rather on the graph structure.
Hence, encoding a graph with an adjacency matrix may be problematic
because a permutation of the rows and columns will result in
the same graph but a different encoding.
An RL agent will not necessarily treat both equally,
even though
all information about the graph is represented.

The intersection between RL and graph drawing has been relatively small so far.
To the best of our knowledge, there is only one unpublished preprint:
Safarli, Zhou, and Wang~\cite{DBLP:journals/corr/abs-2011-00748} use
RL to simulate
classic graph-drawing algorithms, such as force-directed layouts and stress majorization.
Their approach uses Q-learning~\cite{WatkinsD92} where, for every vertex, an agent with its own Q-table is created.
At each iteration, the agent gets an observation where the Q-table reflects the potential reward associated with each action under the current observation: staying or moving in one of the eight cardinal and intercardinal directions.
The agent either picks the action with the highest potential reward or a random action to facilitate exploration.
The reward is positive if the chosen action reduces the energy or stress in the drawing
and negative otherwise.
After each action, the specific Q-table
is updated.
The resulting drawings reach a similar quality as the drawings generated by the corresponding classic algorithms.

\section{Description of Our Approach}
\label{sec:approach}

We use an iterative RL approach in which a \emph{state} is a straight-line drawing of a graph on an integer grid.
An RL agent performs actions that may change the state.
Each action selects one vertex from a shortlist of candidates (described below) and moves it on the grid.
After each action, the agent receives a reward derived from the change in an objective function: either the global crossing number~$\overline{\mathrm{cr}}$ or the local crossing number~$\overline{\mathrm{lcr}}$.
We train one agent for each objective and refer to the resulting algorithms as \textsf{RL(GC)} (global crossings) and \textsf{RL(LC)} (local crossings).
The goal of an agent is to reduce the objective via a sequence of moves; the agent has full information and no adversary.
Following previous work~\cite{DBLP:journals/cj/BekosFGHKSS21}, we apply our algorithm as a post-processing step to an initial layout obtained from a standard graph-drawing algorithm (Kamada--Kawai in our experiments, see \cref{sec:experiments}).

The main idea of our current framework is to train the agent as a local repair operator.
Instead constructing a drawing from scratch, or solving the whole problem in one long episode, we repeatedly present the agent with a drawing that is close to a reasonable layout but still contains crossings.
At the beginning of a repair episode, the environment perturbs the current drawing by a small number of random moves and then asks the agent to improve the resulting drawing within a short horizon.
Thus, the agent learns how to repair local crossing configurations.
We restart the repair process several times from the best drawing found so far, apply small random perturbations, and keep the best drawing over all restarts.

The repair operator is graph-size invariant.
In each state, the agent is presented with a shortlist of candidate vertices and chooses both the vertex to move and the movement to perform.
We use eight directions but only allow to choose from several discrete movement distances.
Further, we give no option of staying because we have the impression that this slows down the process of altering the drawing too much.
The algorithm is realized on an integer grid;
vertex positions are clipped to the drawing area and locally repaired if a move produces an invalid placement.
The agent receives observation data corresponding to the current shortlist of candidate vertices.
This design choice has the purpose of treating vertices that are at structurally similar positions in the graph and the drawing similarly.
We use a neural network as an agent that is trained on large data sets, as retrieving knowledge from previous training phases is a core feature of RL\@.

This viewpoint resembles the approach of Safarli, Zhou, and Wang~\cite{DBLP:journals/corr/abs-2011-00748},
who let each agent ``sit'' on a vertex~$v$ and ask for a direction to move~$v$.
Different from their approach, we do not use a separate agent at every vertex; instead we use a single agent that takes the viewpoint of any shortlisted candidate vertex.
Moreover, they
do not train their agents on a data set but rather fill their tables ad hoc when provided with the input graph.

\subparagraph*{Action Space and Movement of Vertices.}

In each step, the agent chooses a compound action.
The first component selects one entry from the current vertex shortlist, the second component selects one of eight movement directions, and the third component selects one of several discrete movement distances.
The latter are given as powers of $2$; in the main configurations we use the six distances $2^0,2^1,\ldots,2^5$.
Thus, the action space is a product of shortlist slot, direction, and distance scale.
After applying the move on the integer grid, the position is clipped to the drawing area.
If necessary (because the vertex would end up on a different vertex or an edge),
the environment applies local repair logic to obtain a valid vertex position.
This is done by trying the next integer positions in a spiral around the original target position.
We maintain a crossing data structure that is then updated incrementally for the affected incident edges.

\subparagraph*{Vertex Selection.}

We next describe how we construct the shortlist of candidate vertices from which the agent chooses one vertex to move.
We define and score the possible candidates, and the four vertices with the highest score are passed as possible options to the agent.
This shortlist of candidates is recomputed from the current drawing, and the agent selects and moves a single vertex from this shortlist.
Invalid entries on the shortlist, which can occur if fewer than the configured number of candidates (here, four) are available, are masked in the policy output.

For global crossing optimization, the candidate set consists of all vertices incident to at least one edge that is currently crossed.
Each candidate~$v$ is scored by the total number of crossings on its incident edges, divided by a visit-penalty term.
The visit-penalty term starts at~1 and increases by 0.5 for each time that vertex has been selected before.
Intuitively, this focuses the policy on vertices whose movement can affect many current crossings while discouraging repeatedly moving the same vertex.

When optimizing local crossings, let $k$ be the maximum number of crossings on any edge in the current drawing,
and let $E^\star$ be the set of edges with exactly $k$ crossings, also referred to as critical edges.
The candidate set consists of the endpoints of edges in~$E^\star$ and the endpoints of all edges crossing an edge in~$E^\star$.
For scoring every candidate~$v$, the total crossing mass\footnote{%
The crossing mass of a vertex is the sum of the crossing counts of its incident edges.}
of incident edges, the maximum crossing count of an incident edge,
the number of incident critical edges, the number of incident edges that cross a critical edge, and
the distance to the nearest crossing point involving a critical edge contribute positively.
This is divided by the square root of the vertex degree times the same visit-penalty term as above.
The shortlist again contains the highest-scoring vertices.
Intuitively, this focuses the policy on vertices close to the currently critical local crossing structure while still preferring lower-degree and less frequently visited vertices.

Since the agent receives a shortlist of four candidates,
it additionally receives some information about those candidates before making a decision
in the observation vector we describe next.

\begin{figure}
    \centering
    \includegraphics[width=0.95\linewidth]{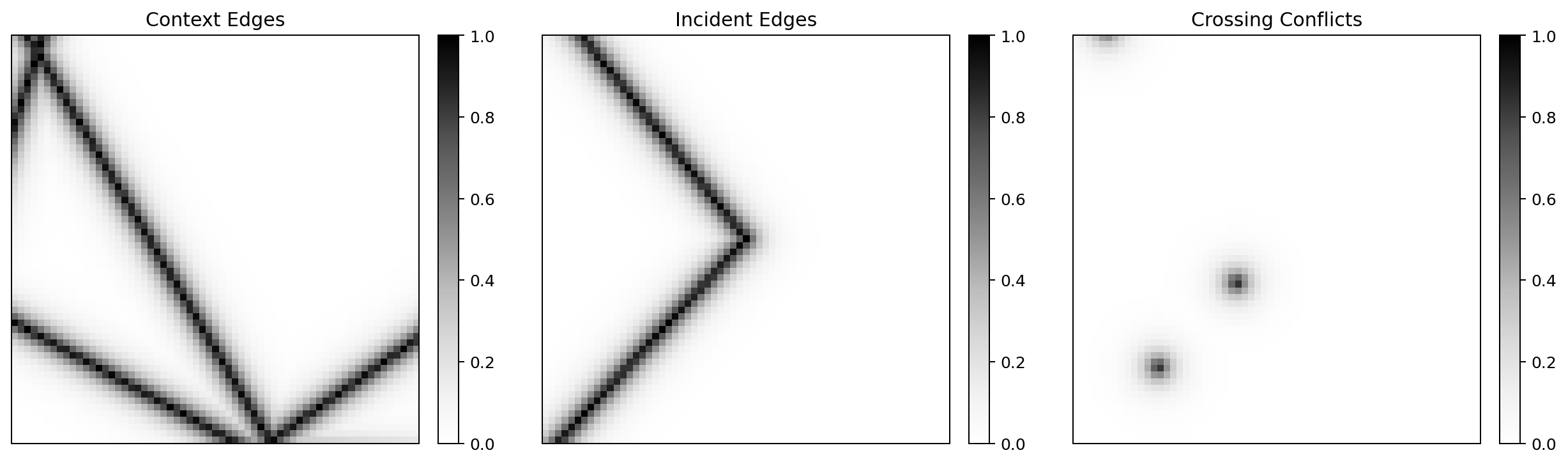}
    \caption{Example of the local pixel map around a candidate vertex.
        The three channels encode nearby non-incident edges, incident edges, and crossing points involving incident edges.}
    \label{fig:pixel_map_greys}
\end{figure}

\subparagraph*{Observation vector.}
The observation vector for a candidate vertex is three-fold:
the first part contains information about the current graph drawing including the local neighborhood of the candidate vertex.
The second part contains additional per-candidate features used by the shortlist mechanism,
such as the heuristic score used for selecting the candidate into the shortlist, degree, visit count,
and the incident crossing mass.
The third part contains a 3-channel image-like patch
of the local neighborhood with the currently considered vertex in the center (see \cref{fig:pixel_map_greys}).

\begin{figure}[t]
    \centering
    \begin{minipage}{.7 \linewidth}
        \newcolumntype{f}{>{\scriptsize}l}
        \newcolumntype{t}{>{\scriptsize}r}
        \setlength{\tabcolsep}{5pt}
        \begin{tabular}{f| t t t t t t t t }
            octant & \multicolumn{1}{t}{I} & II & III & IV & V & VI & VII & VIII \\         
            \hline
            vertices/all & 0.31 & 0.16 & 0.00 & 0.04 & 0.13 & 0.07 & 0.07 & 0.21 \\
            vertices/max & 1.00 & 0.50 & 0.00 & 0.14 & 0.43 & 0.25 & 0.21 & 0.68 \\
            closest ngb.\ & 25.00 & 9.90 & 0.00 & 0.00 & 8.04 & 21.1 & 16.00 & 7.72  \\
            closest non-ngb.\ & 16.42 & 6.28 & 0.00 & 31.6 & 18.36 & 21.00 & 21.00 & 11.80 \\
            ray distance  & 3.55 & 5.07 & 6.00 & 8.49 & 13.00 & 9.19 & 13.00 & 4.93 \\
            crossings & 0.10 & 0.20 & 0.00 & 0.00 & 0.10 & 0.30 & 0.00 & 1.00 \\
            local cr.\ & 0.25 & 0.50 & 0.00 & 0.00 & 0.25 & 0.50 & 0.00 & 1.00
        \end{tabular}
    \end{minipage}
    \hfill
    \begin{minipage}{.29 \linewidth}
        \includegraphics{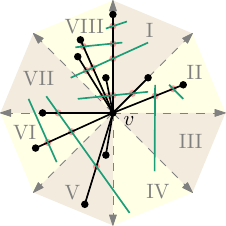}
    \end{minipage}
    \caption{Local view of a vertex~$v$ to its neighborhood in the drawing:
        $v$ is the center of the eight octants
        (octants are represented as half-open angle intervals, including their starting boundary ray).
        In each octant, parameters such as vertex distribution, distances to nearby objects, and incident crossing summaries are measured.
        The table gives an illustrative example of the octant features and normalization;
        the values in the ray-distance row are approximate estimates from the drawing.
        In this example, octant VIII has the largest vertex count and crossing summaries, so the corresponding maximum-normalized entries are~1.}
    \label{fig:observation-space}
\end{figure}

The local neighborhood of a vertex~$v$ in the graph drawing is described with
respect to the eight octants delimited by the eight octagonal rays
going up, up-right, right, down-right, etc.; see \cref{fig:observation-space}.
For each octant, the observation vector encodes
\begin{inparaenum}[(i)]
    \item the number of vertices in the octant normalized by the number of vertices,
    \item the number of vertices in the octant normalized by the maximum over all octants,
    \item the distance to the closest neighbor in the octant,
    \item the distance to the closest non-neighbor in the octant,
    \item the distance to the first non-incident edge hit by the left boundary ray of the octant,
    \item the normalized sum of crossing counts over incident edges of~$v$ in the octant, and
    \item the normalized maximum crossing count of an incident edge of~$v$ in the octant.
\end{inparaenum}

The first two entries describe the distribution of all other vertices around~$v$,
using two normalizations: by the total number of other vertices and by the largest octant count.
The neighbor relation is represented by the closest-neighbor distance in each octant.
The octant-related measures are normalized where appropriate:
vertex-count features are normalized either by the total number of other vertices or by the maximum octant count,
and crossing-summary features are normalized by their maximum over the eight octants.
Distances are kept as absolute grid distances, with empty octants represented by zero for vertex distances
and by the ray length if no non-incident edge is hit.

Together, the five geometric octant features yield a vector of length~40 for the local neighborhood;
the two crossing summaries yield two additional vectors of length~8.
To make the behavior of the agent invariant for rotating the coordinate system,
whenever we call the agent, we rotate the octant entries in the observation vector
such that the direction with the largest incident crossing mass is aligned with octant~I.
The current global and local crossing numbers of the full drawing are added as scalar features.

For each candidate vertex, we additionally store metadata consisting of the normalized shortlist score,
normalized degree, normalized visit count, normalized incident crossing mass, and a validity indicator.
A separate shortlist mask marks which shortlist slots are valid.

In addition to the octant features, the implementation uses a local pixel map around each candidate vertex;
see \cref{fig:pixel_map_greys}.
For every candidate vertex on the shortlist, this pixel map has three ``grayscale'' channels on a fixed $63 \times 63$ patch centered at the vertex:
one channel encodes nearby non-incident edges,
one channel encodes incident edges,
and one channel encodes nearby crossing points involving incident edges of the candidate vertex.
Internally, each channel of the pixel map is a vector of size $63^2 = 3969$
whose entries are grayscale values:
larger values indicate pixels closer to the respective edge or crossing objects, and smaller values indicate pixels farther away from them.
The pixel map is processed by a convolutional neural network into a 64-dimensional representation, which is combined with the non-pixel features of the candidate vertex.
The pixel map is rotated consistently with the rotation of the observation vector.

The full observation vector in each iteration is obtained by stacking these features for all candidate vertices on the shortlist.

\subparagraph*{Reward Function.}
\label{sec:reward-function}

The reward function depends on the objective.
The following equations describe the immediate metric-change reward before adding the best-so-far bonus described below.
For the global crossing number, the reward function is based on the number of removed crossings:
\begin{equation*}
    \small
    \textrm{reward}_\textrm{GC} = \begin{cases}
        \overline{\mathrm{cr}}(\Gamma) - \overline{\mathrm{cr}}(\Gamma') &\text{, if } \overline{\mathrm{cr}}(\Gamma) \ne \overline{\mathrm{cr}}(\Gamma') \\
        -0.01 &\text{, otherwise.}
    \end{cases}
\end{equation*}
Here, $\Gamma$ is the drawing before and $\Gamma'$ is the drawing after performing an action.
If the action leaves all reward-relevant metrics unchanged, we give a small negative reward to discourage metric-neutral moves.

The reward for the local crossing number combines the change in the local and global crossing number,
where $m^*(\Gamma)$ is the number of edges having $\overline{\mathrm{lcr}}(\Gamma)$ crossings on it:
\begin{equation*}
    \small
    \textrm{reward}_\textrm{LC} = \begin{cases}
        10 \left(\overline{\mathrm{lcr}}(\Gamma) - \overline{\mathrm{lcr}}(\Gamma')\right)
        + 0.05\left(\overline{\mathrm{cr}}(\Gamma) - \overline{\mathrm{cr}}(\Gamma')\right)
        & \text{, if } \overline{\mathrm{lcr}}(\Gamma) \ne \overline{\mathrm{lcr}}(\Gamma')\\
        0.1 \left(m^*(\Gamma) - m^*(\Gamma')\right)
        + 0.05\left(\overline{\mathrm{cr}}(\Gamma) - \overline{\mathrm{cr}}(\Gamma')\right)
        & \text{, if } m^*(\Gamma) \ne m^*(\Gamma')
        \text{ or } \overline{\mathrm{cr}}(\Gamma) \ne \overline{\mathrm{cr}}(\Gamma') \\
        -0.01 & \text{, otherwise.}
    \end{cases}
\end{equation*}

The idea is to primarily reward reducing the local crossing number, but also give a reward if the agent moves the layout in the right direction.
Thus, we also reward reducing the number of edges with $\overline{\mathrm{lcr}}(\Gamma)$ crossings if the local crossing number did not change,
and include the change in global crossing number in both non-sparse local-reward cases.

Since we regularly reset the layout to previous states, we additionally give a bonus of \(5\cdot\Delta^\star\), where \(\Delta^\star\) is the improvement over the best objective value found so far.
In the global setting, \(\Delta^\star\) is the improvement in the best global crossing number; in the local setting, it is the improvement in the best local crossing number.
This explicitly rewards finding an overall improved drawing during the rollout, even if the final drawing of the episode is not the best.

\subparagraph*{Policy Network, Training, and Evaluation.}

Our algorithm starts with employing a classical graph-drawing algorithm
to obtain a baseline drawing.
During training, each graph is treated as an episodic repair environment initialized from a slightly perturbed drawing derived from it.
The policy is represented by a neural network that processes each candidate vertex individually: the spatial pixel map of each vertex is encoded by a convolutional neural network, its tabular features are encoded by a multilayer perceptron (MLP), and the resulting representations are concatenated and passed through a fusion MLP to produce a single embedding. 
The output is a masked product distribution over shortlist slot, movement direction, and movement distance.

We use proximal policy optimization (PPO)~\cite{DBLP:journals/corr/SchulmanWDRK17} to train the network.
During training, actions are sampled from the policy distribution rather than chosen by an \(\epsilon\)-greedy rule, and PPO updates the policy from the rewards obtained after the repair moves.
We always save the best drawing.
At evaluation time, we use several restarts from the currently best drawing, each with small perturbations, and return the best drawing found over all restarts.

\section{Experimental Evaluation}
\label{sec:experiments}

We are interested in answering the following research questions:
\begin{itemize}[widest={RQ2}, leftmargin=*]
    \item[\bfseries \textsf{\textcolor{blue!60!black}{RQ1}}]\label{rq:1} How well does a reinforcement-learning approach perform in computing drawings with a low global or local crossing number? 
    \item[\bfseries \textsf{\textcolor{blue!60!black}{RQ2}}]\label{rq:2} How do existing global crossing minimization algorithms perform in finding drawings with low local crossing number? 
\end{itemize}

\subparagraph*{Implementation details.}

The implementation is written in Python~3.12, with performance-critical routines implemented in C++ and exposed to Python via \texttt{pybind11}.
These C++ routines handle crossing computations, batched computation of octant observations, and pixel maps.
The custom environment is implemented on top of
\texttt{Gymnasium} (v\texttt{1.1.1}).
We train the policy with maskable PPO using \texttt{stable-baselines3} (v\texttt{2.6.0}) and \texttt{sb3-contrib} (v\texttt{2.4.0});
the neural network itself is implemented in \texttt{PyTorch} (v\texttt{2.6.0}).
Training was done on an H200 GPU, and the CPU was an AMD EPYC 9555 Turin with 32 cores allocated. 

Our models \textsf{RL(GC)} and \textsf{RL(LC)}
use a shortlist of four vertices,
a repair horizon of 32 steps, six movement distances,
and a grid of size $1000 \times 1000$.
Training uses 24 vectorized environments wrapped in \texttt{VecMonitor}.
Each model is trained for \(5\cdot 10^6\) PPO steps, with a rollout length of 256, a batch size of 128, and 8 PPO epochs per rollout.
We use \(\gamma=0.99\), GAE parameter \(\lambda=0.99\), and linearly decay both the learning rate and entropy coefficient during training.
At evaluation time, we use 16
restarts with two perturbation moves each.

\subparagraph*{Benchmark Graphs.}

We use two classes of graphs. 
First, the \rome graphs\footnote{\url{http://graphdrawing.org/data.html}}, 
commonly used in evaluating graph drawing tasks~\cite{DBLP:journals/cgf/BartolomeoCSPWD24,di1997experimental}.
These are 11,534 relatively sparse graphs with 10--100 vertices and 9--158 edges.
Second, 
we used \texttt{networkx}~\cite{hagberg2008exploring} to generate random graphs; namely, extended Barabási-Albert~(\BA) graphs~\cite{albert2000topology} with 50--150 vertices, preferential attachment parameter $m\in \{1,2,3\}$,
and reconfiguration parameters $p\in [0,0.1], q\in[0,0.2]$.

We filter 
both sets to retain only connected, non-planar instances 
and then recursively remove all degree-1 vertices from the resulting graphs.
After obtaining a drawing with low (local) crossing number, degree-1 vertices could be reinserted in an $\epsilon$-neighborhood of the adjacent vertex without introducing crossings (ignoring the grid).
This preprocessing is applied uniformly before calling all algorithms and mirrors the simplifications used by the \textsf{Vertex Movement} and \textsf{Edge Insertion} implementations.
After this filtering step, the final \rome set contains 8,252 graphs with 9--91 vertices and 13--148 edges;
the mean size is 48.27 vertices and 69.91 edges.
The final \BA set contains 1,500 graphs with 13--150 vertices and 19--495 edges;
the mean size is 88.82 vertices and 222.05 edges.
For the \BA graphs, the filtering removes most graphs with parameter $m=1$, as these are mostly planar.
Finally, an initial drawing used as input to the algorithms is obtained with the
Kamada--Kawai algorithm~\cite{kamada1989algorithm}, 
which uses spring forces between vertices proportional to their graph-theoretic distance. 
In initial experiments, it created drawings with usually fewer global and local crossings than the Fruchterman--Reingold algorithm~\cite{fruchterman1991graph}.
The \rome graphs are split into training (80\%) and testing (20\%) instances.
Similarly, 1,000 \BA graphs were used for training and 500 for testing.

\subparagraph*{Benchmark Algorithms.}

As generic baselines,
we use
the algorithms
\textsf{Kamada-Kawai}~(\textsf{KK}) and
\textsf{Fruchterman-Reingold}~(\textsf{FR}), both implemented in \texttt{networkx}.
\textsf{KK}
also produces the initial layout used as input to the other algorithms. 
We also compare with another machine-learning-based model, namely the \emph{GAN-based framework} \textsf{SmartGD}~\cite{wang2023smartgd}.
In particular, the authors released a model trained to minimize the crossing number, which we use in our evaluation.\footnote{\url{https://github.com/yolandalalala/SmartGD}}
It was trained on the \rome graphs, like our algorithm partially was.
Regarding other heuristics, we compare to the \emph{stochastic gradient descent} method \textsf{(SGD)\textsuperscript{2}}~\cite{ahmed2022multicriteria},\footnote{\url{https://github.com/tiga1231/graph-drawing/tree/sgd}}
the \emph{geometric local-optimization} approaches
\textsf{Vertex Movement}~(\textsf{VM}) and \textsf{Edge Insertion}~(\textsf{EI})~\cite{radermacher2019geometric},\footnote{Source code obtained from the authors.}
and the \emph{probabilistic hill-climbing} method \textsf{Tübingen-Bus}~(\textsf{TB}).\footnote{While \textsf{TB} was originally designed for
upward drawings, we could easily adjust it to drop the upward constraint~\cite{maxi}. Source code obtained from and adjusted according to the instructions of the authors.}.
We use the default parameters given in the implementations or papers;
see \cref{sec:preliminaries} for details on the algorithms.

\subparagraph*{Results.}
We run all algorithms on the test sets, recording global and local crossing numbers and running time with a time limit of 900 seconds per graph and algorithm.
Evaluations were run on one AMD EPYC 9555 Turin CPU with 128GB of memory; \textsf{SmartGD}, \textsf{(SGD)\textsuperscript{2}}, and \textsf{KK}/\textsf{FR} are Python-based, while the other baselines are implemented in C++.
We first report the performance results for \hyperref[rq:1]{\bfseries \textsf{\textcolor{blue!60!black}{\emph{RQ1}}}} and then compare global and local crossing-number rankings of the non-RL algorithms for \hyperref[rq:2]{\bfseries \textsf{\textcolor{blue!60!black}{\emph{RQ2}}}}.

\begin{figure}[p]
    \begin{minipage}[b]{\textwidth}
        \centering
        \begin{subfigure}[b]{0.48\textwidth}
            \centering
            \includegraphics[scale=.37]{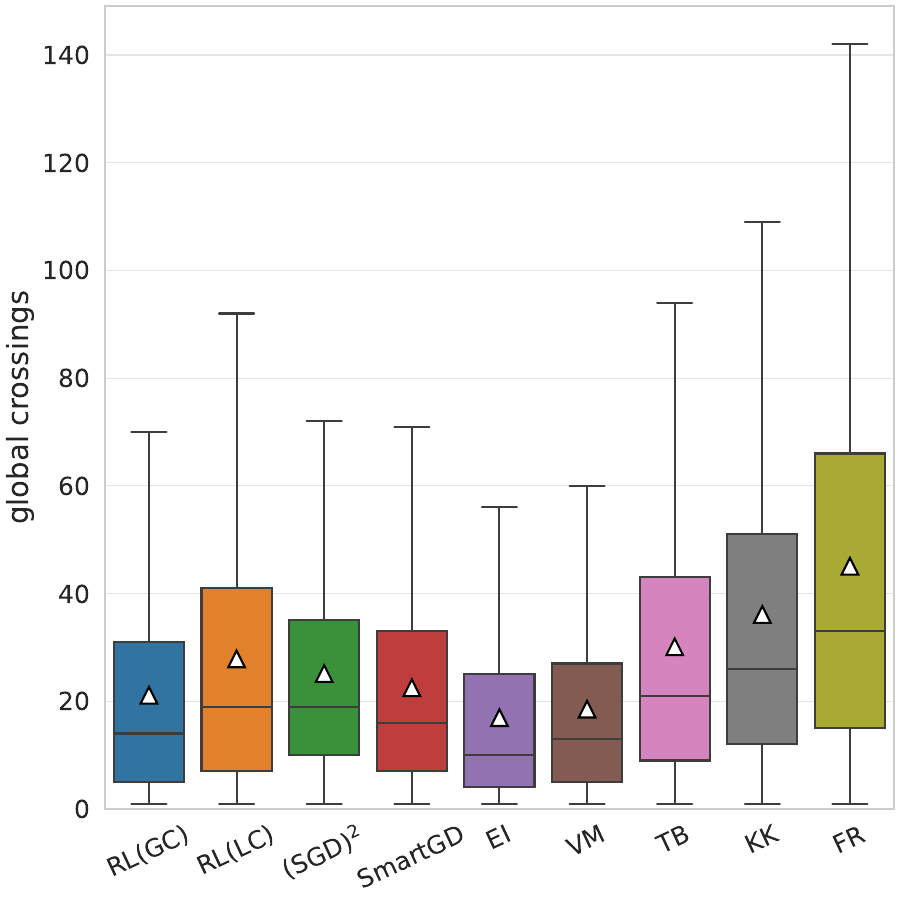}
        \end{subfigure}
        \hfil
        \begin{subfigure}[b]{0.48\textwidth}
            \centering
            \includegraphics[scale=.37]{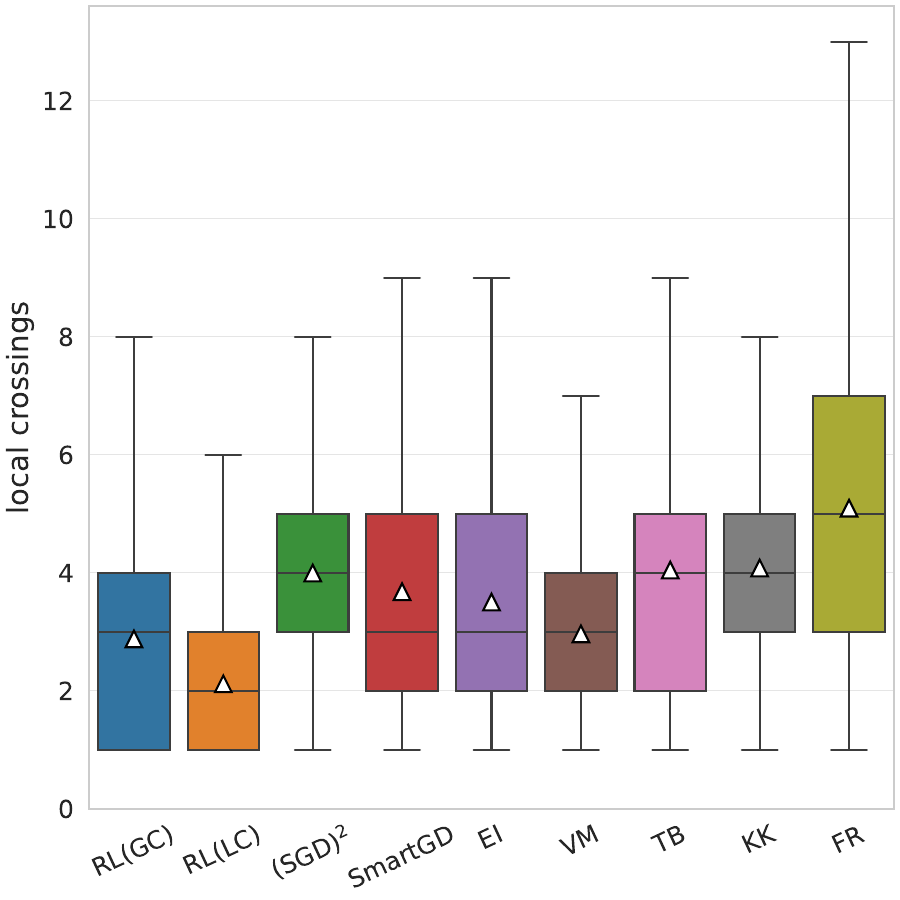}
        \end{subfigure}
        \caption{Box plots of the global (left) and local (right) crossing number on the \rome graphs.
        The thin bars indicate the full range of crossing numbers
        over all generated drawings of the data set,
        while the thick bars indicate the two middle quartiles
        with the horizontal line segment indicating the median.
        The small triangles indicate the arithmetic mean.}
        \label{fig:rome-boxplots}
    \end{minipage}
    
    \bigskip
    
    \begin{minipage}[b]{\textwidth}
        \centering
        \begin{subfigure}[b]{0.48\textwidth}
            \centering
            \includegraphics[scale=.37]{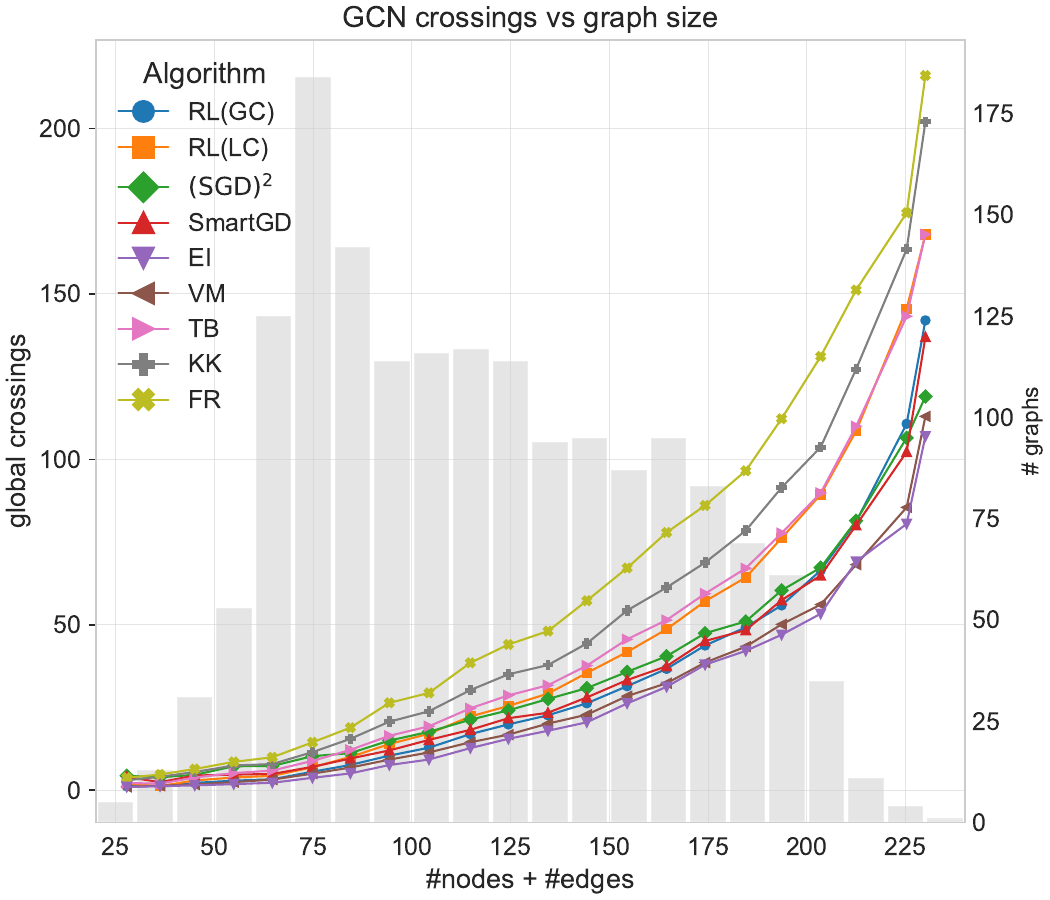}
        \end{subfigure}
        \hfill
        \begin{subfigure}[b]{0.48\textwidth}
            \centering
            \includegraphics[scale=.37]{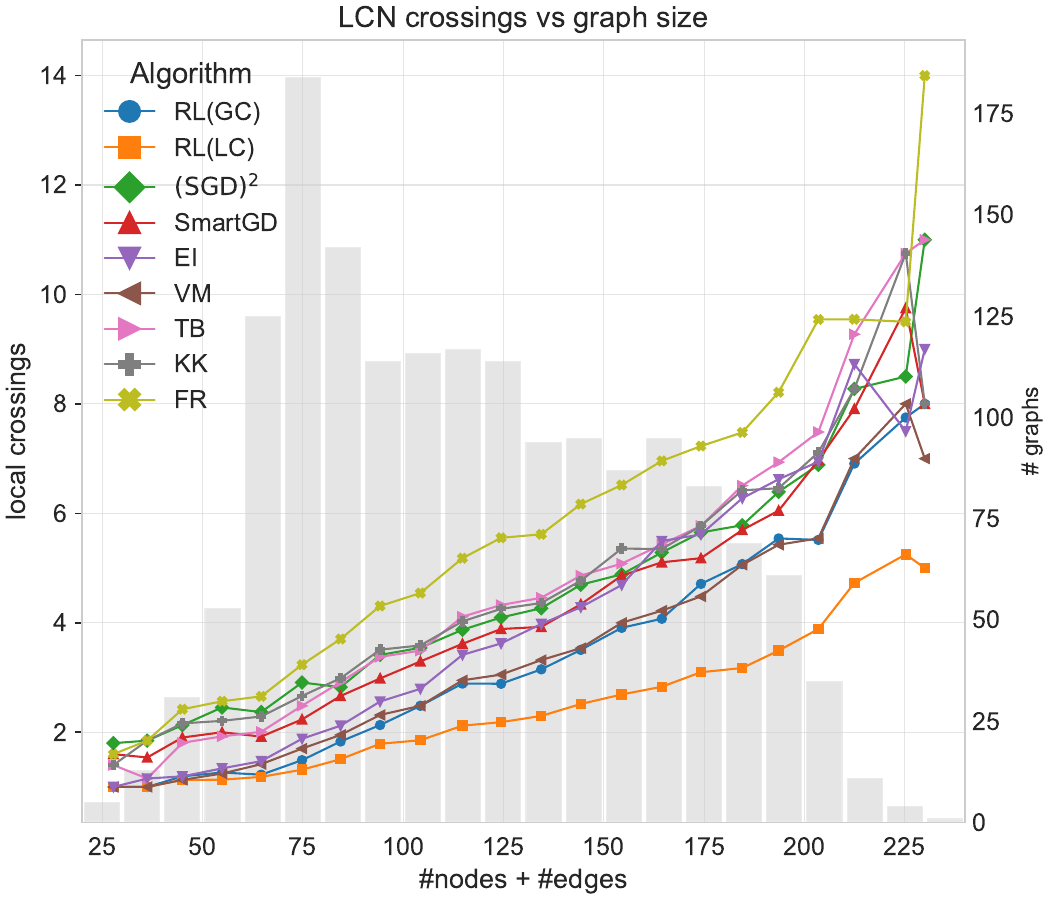}
        \end{subfigure}
        \caption{Average global (left) and local (right) crossing numbers
            grouped by graph sizes on the \rome graphs.
            Each dot represents a bucket of graphs
            whose number of vertices plus edges lies in a range of 10.
            The number of graphs in a bucket
            is indicated by the gray bars in the background.}
        \label{fig:rome-trend-curve}
    \end{minipage}
    
    \bigskip
    
    \begin{minipage}[b]{\textwidth}
        \centering
        \includegraphics[width=0.9\linewidth]{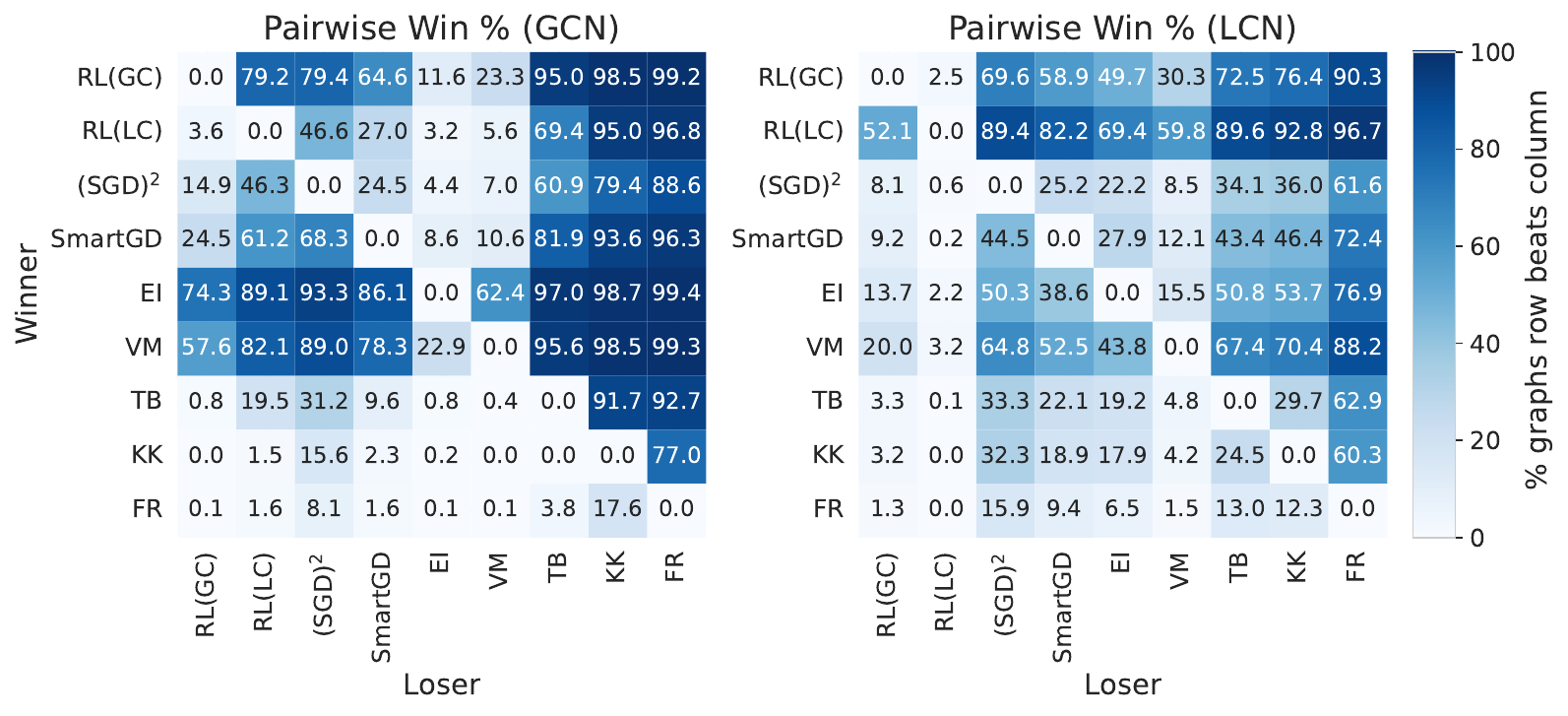}
        \caption{Pairwise win percentage on the \rome graphs.}
        \label{fig:rome-pairwise}
    \end{minipage}
\end{figure}

On the \rome graphs (\cref{fig:rome-boxplots,fig:rome-trend-curve}),
\textsf{EI} and \textsf{VM} obtain the lowest mean global crossing numbers, 16.95 and 18.52, followed by \textsf{RL(GC)} and \textsf{SmartGD} with means 21.11 and 22.49.
For local crossing number, \textsf{RL(LC)} performs best with mean 2.12 and median 2, followed very closely by \textsf{RL(GC)} and \textsf{VM}, both with median 3.
Pairwise win percentages are shown in \cref{fig:rome-pairwise}.
We remark that for many \rome graphs,
the exact global crossing numbers are known~\cite{DBLP:conf/esa/ChimaniW16,crossing-numbers}.
However, these are the topological crossing numbers that are not necessarily
equal to the rectilinear crossing numbers,
which may be higher.
So they can only be seen as lower bounds.
Still, we provide an overview on how far the
drawings generated by the tested algorithms
are away from these exact global crossing numbers in \cref{tab:comparison-rome-cr}.

\begin{table}[tb]
    \centering
    \begin{tabular}{lrrrrrr}
        \toprule
        Algorithm & \makecell{mean} & \makecell{median} & \makecell{gap mean (\%)} & \makecell{gap median (\%)} & \makecell{exact matches} & \makecell{\% graphs} \\
        \midrule
        \textsf{RL(GC)} & 17.72 & 13 & 126.0 & 114.3 & 122 & 8.1 \\
        \textsf{RL(LC)} & 22.19 & 17 & 183.8 & 171.4 & 91 & 6.1 \\
        \textsf{(SGD)\textsuperscript{2}} & 20.76 & 17 & 240.6 & 171.4 & 15 & 1.0 \\
        \textsf{SmartGD} & 18.01 & 14 & 161.1 & 130.8 & 43 & 2.9 \\
        \textsf{EI} & 12.85 & 9 & 47.1 & 44.4 & 389 & 26.0 \\
        \textsf{VM} & 14.52 & 11 & 90.1 & 80.0 & 150 & 10.0 \\
        \textsf{TB} & 23.98 & 19 & 235.8 & 200.0 & 28 & 1.9 \\
        \textsf{KK} & 29.04 & 23 & 322.1 & 271.4 & 13 & 0.9 \\
        \textsf{FR} & 36.75 & 29 & 441.5 & 371.4 & 7 & 0.5 \\
        \bottomrule
    \end{tabular}
    \caption{Comparison to the (exact global topological) crossing numbers for 1497 (of 1649) \rome graphs,
        where the mean is 7.99 and the median is 6.
        Relative gaps are averaged over the instances.
        Note that we compare here (sometimes strictly larger) rectilinear with topological crossing numbers.}
    \label{tab:comparison-rome-cr}
\end{table}

\begin{figure}[t]
    \centering
    \begin{subfigure}[b]{0.44\textwidth}
        \centering
        \includegraphics[scale=.42,trim=0 0 100 0,clip]{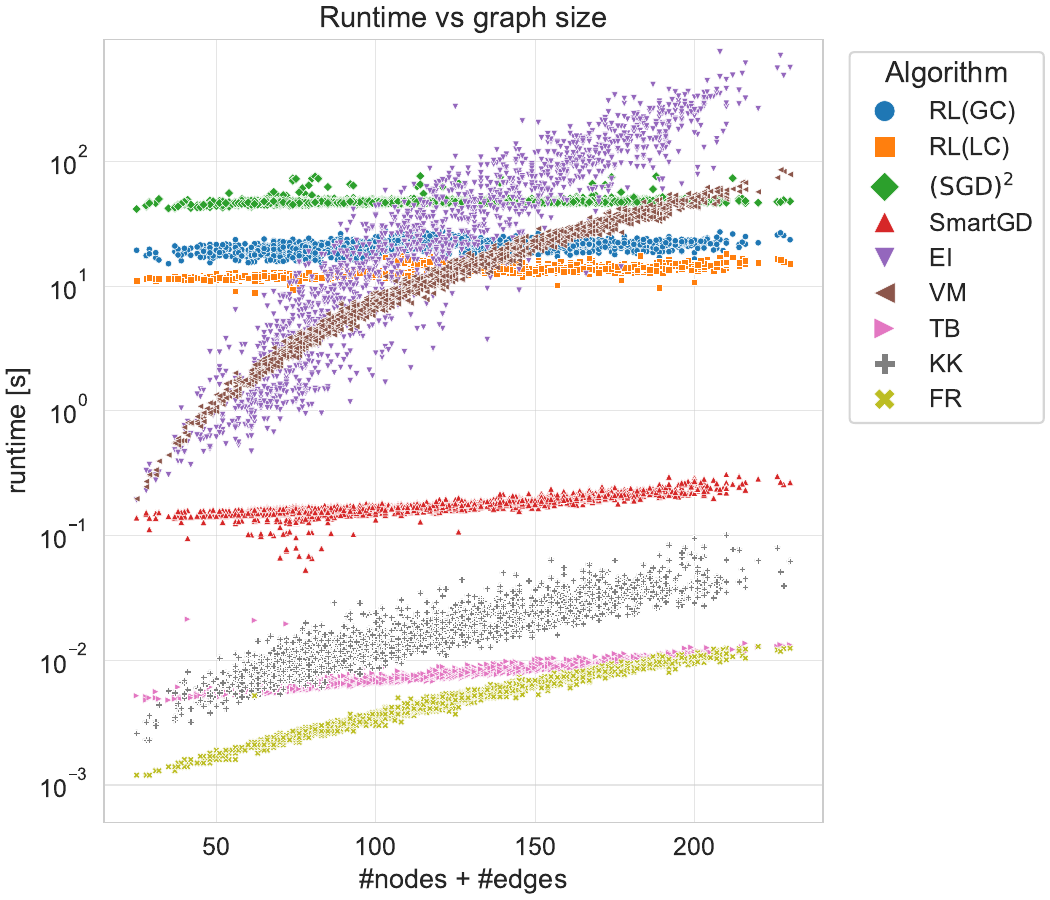}
        \subcaption{\rome graphs}
        \label{fig:rome-runningtimes}
    \end{subfigure}
    \hfill
    \begin{subfigure}[b]{0.52\textwidth}
        \centering
        \includegraphics[scale=.42]{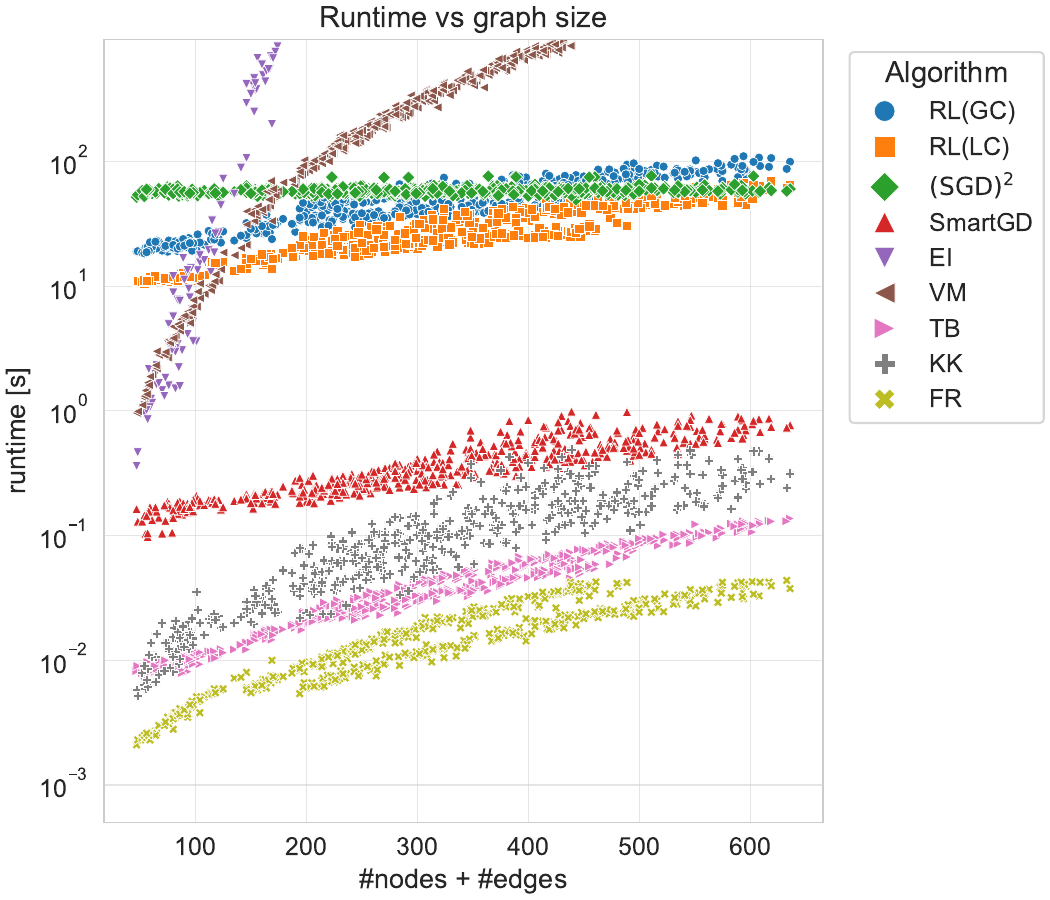}
        \captionsetup{textformat=simple}
        \subcaption{\BA graphs}
        \label{fig:ba-runningtimes}
    \end{subfigure}
    \caption{Scatter plots of the runtimes vs.\ graph sizes. A mark represents a graph.}
    \label{fig:runningtimes}
\end{figure}

On the \rome graphs, \textsf{FR}, \textsf{TB}, \textsf{KK}, and \textsf{SmartGD} have median running times below 0.2 seconds, while \textsf{VM}, \textsf{RL(LC)}, \textsf{RL(GC)}, \textsf{EI}, and \textsf{(SGD)\textsuperscript{2}} have medians of 10.39, 13.09, 21.08, 21.45, and 47.72 seconds, respectively; see \cref{fig:rome-runningtimes}.
Among these slower methods, \textsf{(SGD)\textsuperscript{2}} is the slowest and shows nearly constant running time due to its fixed iteration count.

\begin{figure}[p]
    \begin{minipage}[b]{\textwidth}
        \centering
        \begin{subfigure}[b]{0.48\textwidth}
            \centering
            \includegraphics[scale=.37]{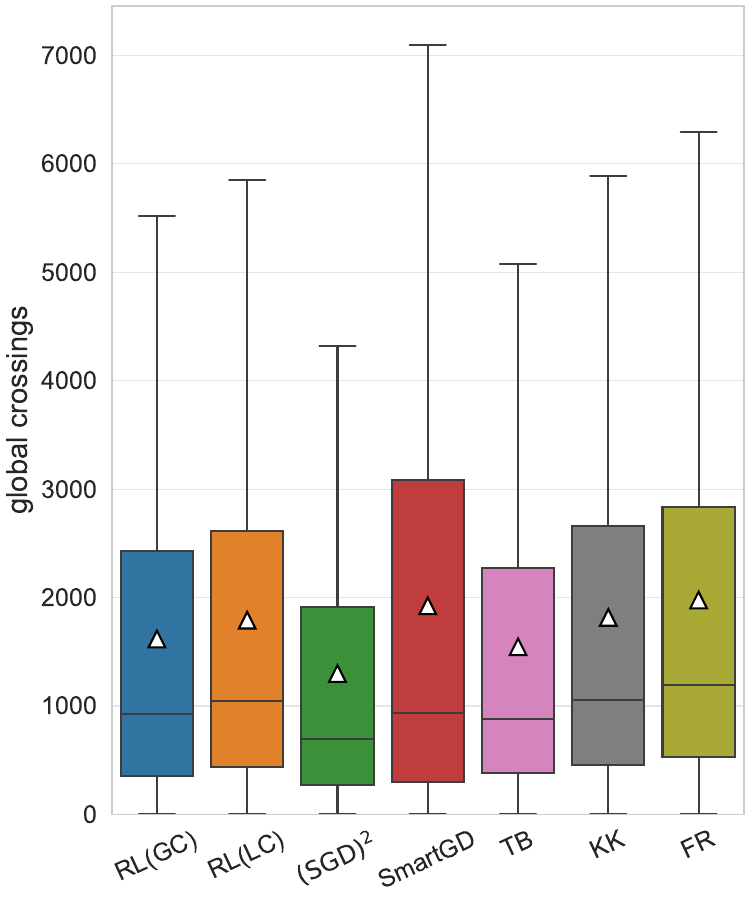}
        \end{subfigure}
        \hfil
        \begin{subfigure}[b]{0.48\textwidth}
            \centering
            \includegraphics[scale=.37]{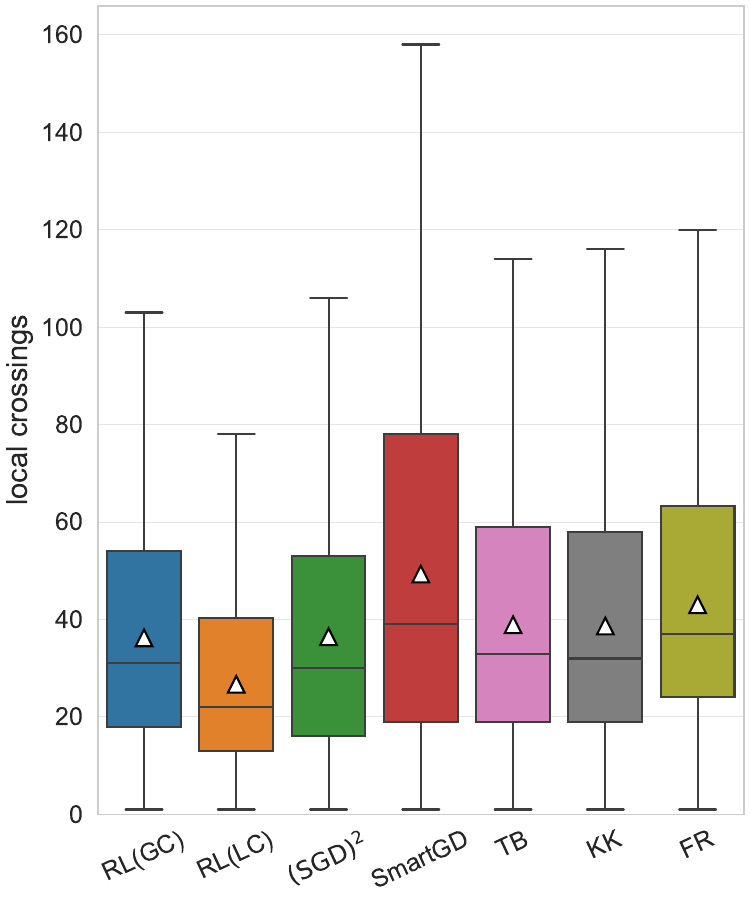}
        \end{subfigure}
        
        \bigskip
        
        \begin{subfigure}[b]{0.48\textwidth}
            \centering
            \includegraphics[scale=.37]{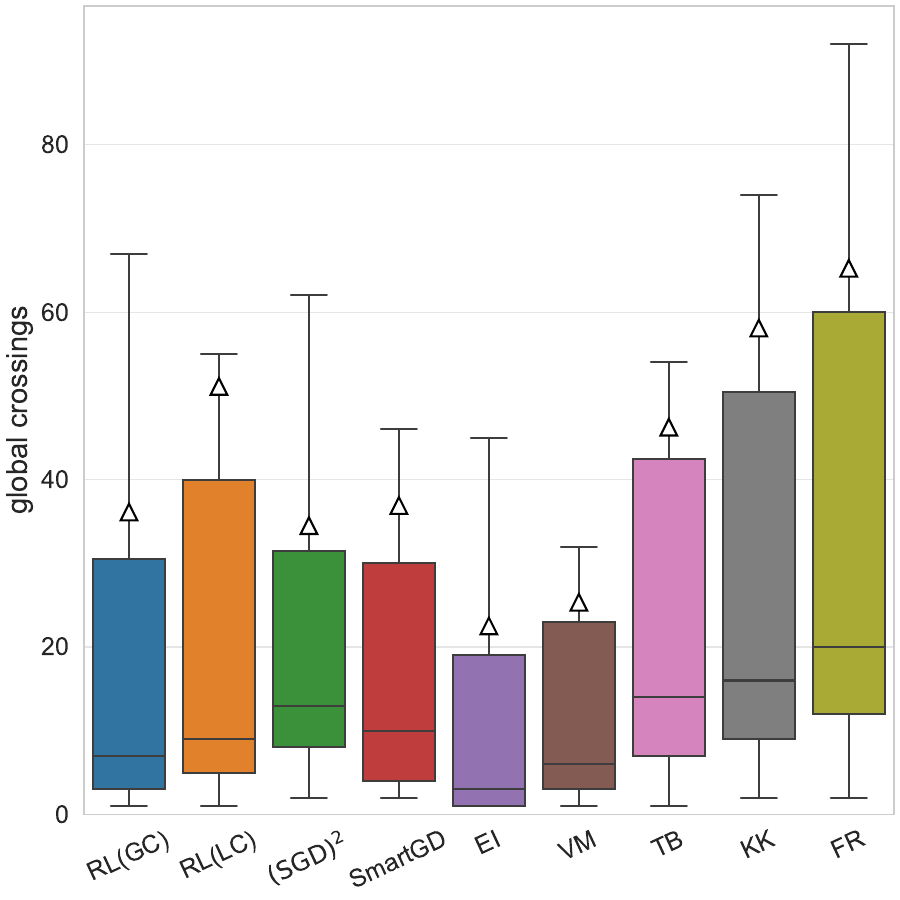}
        \end{subfigure}
        \hfil
        \begin{subfigure}[b]{0.48\textwidth}
            \centering
            \includegraphics[scale=.37]{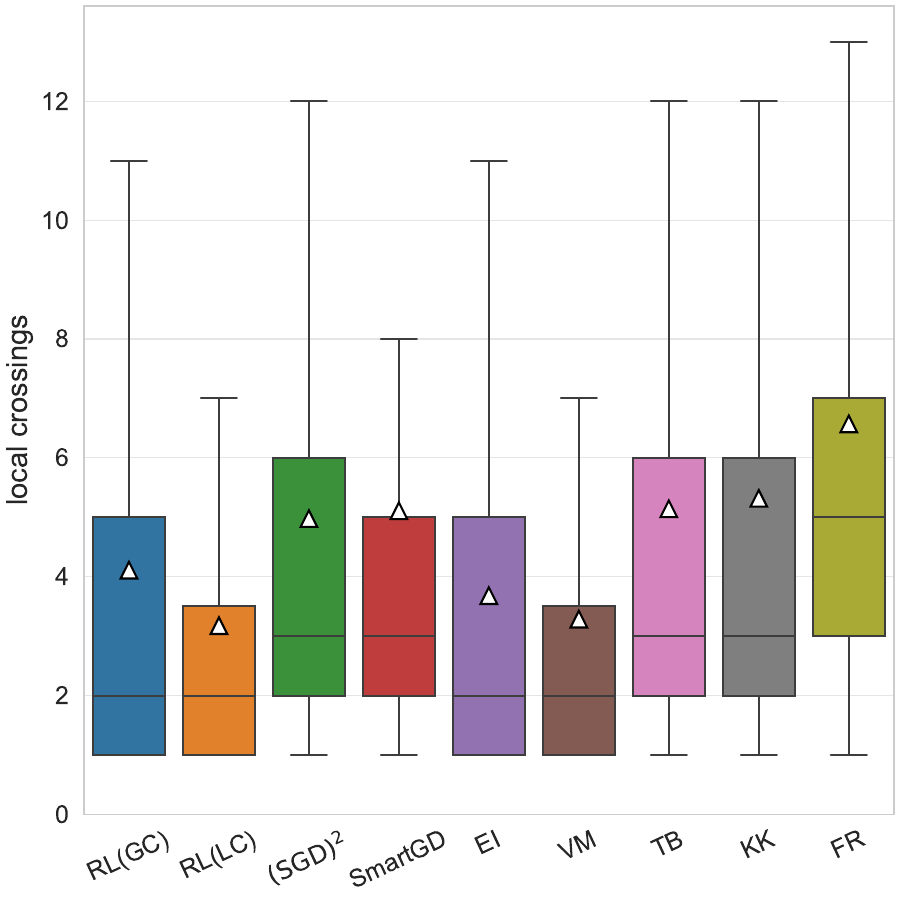}
        \end{subfigure}
        
        \caption{Box plots of the global (left) and local (right) crossing number on the \BA graphs.
            The top row shows the distribution over all instances,
            but excludes the algorithms \textsf{EI} and \textsf{VM},
            which did not finish on all instances.
            The bottom shows the distribution over
            only those instances where all algorithms produced a drawing.
            The thin bars indicate the full range of crossing numbers
            over all generated drawings of the data set,
            while the thick bars indicate the two middle quartiles
            with the horizontal line segment indicating the median.
            The small triangles indicate the arithmetic mean.}
        \label{fig:ba-boxplots-all-instances}
    \end{minipage}
    
    \bigskip
    
    \begin{minipage}[b]{\textwidth}
        \centering
            \begin{subfigure}[b]{0.48\textwidth}
            \centering
            \includegraphics[scale=.4]{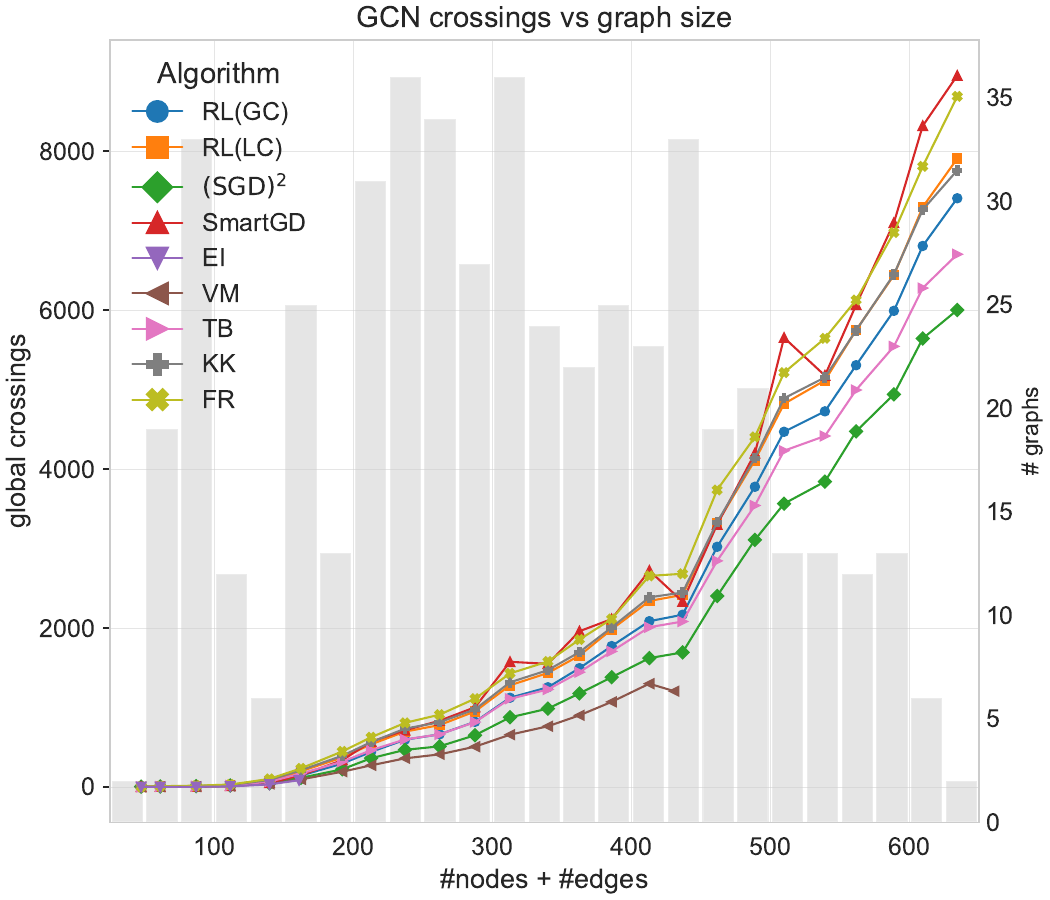}
        \end{subfigure}
        \hfill
        \begin{subfigure}[b]{0.48\textwidth}
            \centering
            \includegraphics[scale=.4]{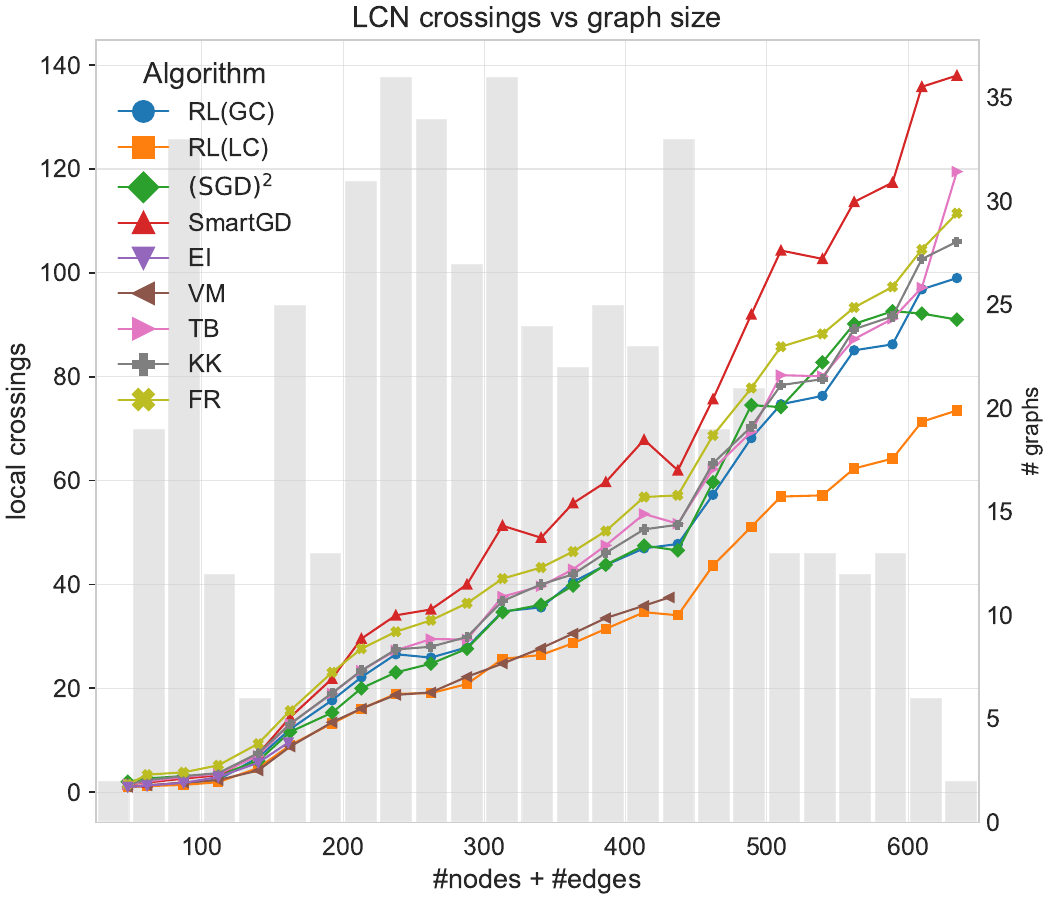}
        \end{subfigure}
        \caption{Average global (left) and local (right) crossing numbers
            grouped by graph sizes on the \BA graphs.
            Each dot represents a bucket of graphs
            whose number of vertices plus edges lies in a range of 25.
            The number of graphs in a bucket
            is indicated by the gray bars in the background.
            Note that the curves of \textsf{EI} and \textsf{VM}
            stop early because these algorithms did not finish
            on the larger graphs.}
        \label{fig:ba-trend-curve}
    \end{minipage}
\end{figure}

\begin{figure}[t]
    \centering
    \includegraphics[width=0.9\linewidth]{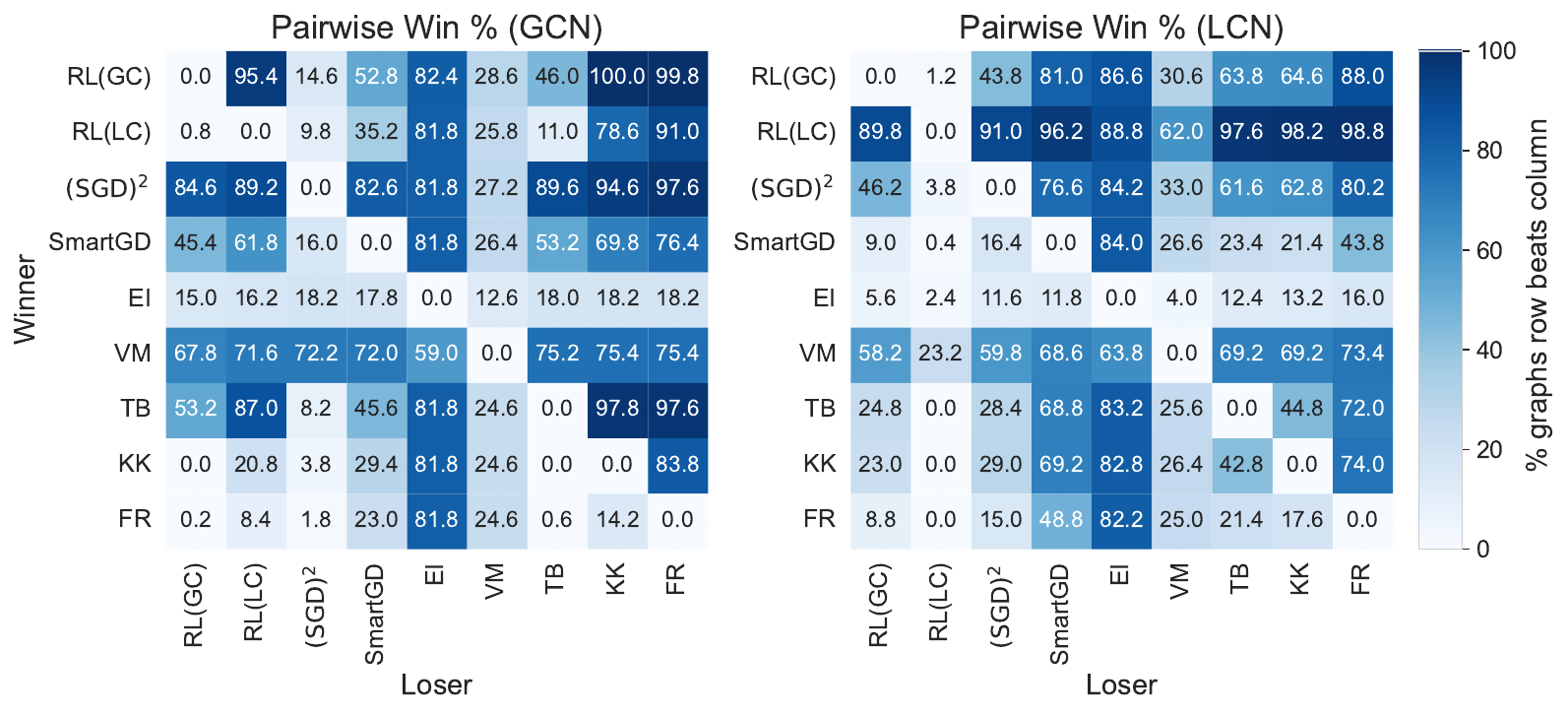}
    \caption{Pairwise win percentage on the \BA graphs.
        If an algorithm did not finish a graph in time, while the other algorithm does, this is counted as a loss for the algorithm that did not finish.}
    \label{fig:ba-pairwise}
\end{figure}

\begin{figure}[p]
    \centering
    \includegraphics[width=0.95\linewidth]{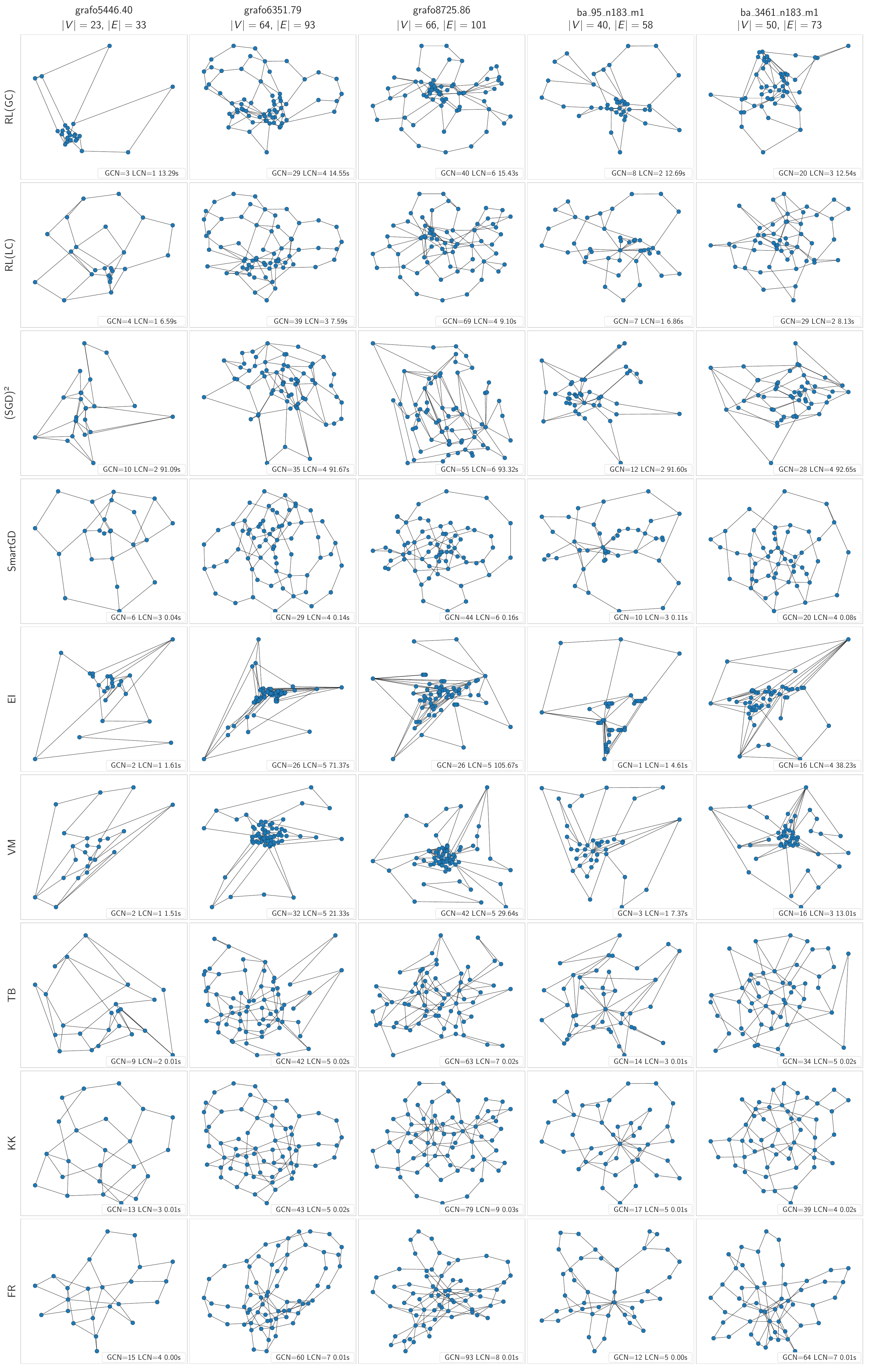}
    \caption{Example drawings obtained from the different algorithms on 5 selected \rome and \BA graphs.
        Included are the global crossing number (GCN) and local crossing number (LCN) of the drawings as well as the running time.}
    \label{fig:sample-outputs}
\end{figure}

For the \BA graphs, the global and local crossing-number results are shown in \cref{fig:ba-boxplots-all-instances,fig:ba-trend-curve}.
Note that \textsf{EI} and \textsf{VM} solved only 91 and 377 instances in time, respectively.
On this restricted set, \textsf{EI} is best for global crossing number, while \textsf{VM} and \textsf{RL(LC)} are also among the strongest methods for local crossing number; however, \textsf{EI} and \textsf{VM} come with severe timeout issues on the full benchmark set.
Among the algorithms that finish all 500 instances, \textsf{(SGD)\textsuperscript{2}} is best for global crossing number with mean 1299.70, followed by \textsf{TB} and \textsf{RL(GC)} with means 1545.72 and 1617.8, while \textsf{RL(LC)} is best for local crossing number with mean 26.69 and median 22.
The running-time trend is similar to the \rome graphs: \textsf{EI} and \textsf{VM} suffer from timeouts, and the median running times of \textsf{RL(LC)}, \textsf{RL(GC)}, and \textsf{(SGD)\textsuperscript{2}} increase to 25.04, 45.07, and 57.91 seconds on \BA; see \cref{fig:ba-runningtimes}.
Pairwise win percentages are given in \cref{fig:ba-pairwise}.
Sample drawings are shown in \cref{fig:sample-outputs}.

\begin{table}[t]
\centering
\caption{Agreement between global and local crossing-number performance for the 7 non-RL algorithms.
Aggregate rank correlation is Spearman's $\rho$ between rankings by mean global crossing number and mean local crossing number.
Median graph-wise rank correlation is the median Spearman's $\rho$ obtained by ranking algorithms separately on each graph.
Pairwise concordance is computed over $N_{\mathrm{pair}}$ strict algorithm-pair comparisons, excluding ties and missing values.}
\label{tab:gcn_lcn_agreement}
\begin{tabular}{lrrrrr}
\toprule
Dataset & $|G|$ & Aggregate $\rho$ & Median graph-wise $\rho$ & $N_{\mathrm{pair}}$ & Pairwise concordance \\
\midrule
Rome    & 1{,}650 & 0.964 & 0.671 & 23{,}986 & 0.812 \\
BA      &   500   & 0.929 & 0.703 &  6{,}719 & 0.796 \\
Rome+BA & 2{,}150 & 0.929 & 0.677 & 30{,}705 & 0.809 \\
\bottomrule
\end{tabular}
\end{table}

For \hyperref[rq:2]{\bfseries \textsf{\textcolor{blue!60!black}{\emph{RQ2}}}}, we compare the global and local crossing-number performance of the non-RL algorithms.
We exclude the RL methods here because the question asks how existing global crossing minimization algorithms behave with respect to local crossing number.
The comparison is made in three complementary ways.

First, we rank the seven algorithms by their mean global crossing number and by their mean local crossing number on each data set.
The resulting rankings are shown in \cref{tab:aggregate_gcn_lcn}.
We then compute Spearman's $\rho$ between these two aggregate rankings.
This measures whether algorithms that perform well on average for global crossing minimization also tend to perform well on average for local crossing minimization.

Second, we repeat the same idea separately for each graph:
for a fixed graph, we rank the algorithms once by global crossing number and once by local crossing number and compute the rank correlation between the two rankings.
The graph-wise correlations are then summarized by their mean and median over all graphs.
These values show whether the aggregate agreement is also visible on typical individual instances.

Third, we compute a pairwise concordance score.
For every graph and every pair of algorithms, we check whether the algorithm with lower global crossing number also has lower local crossing number, ignoring ties and missing values.
This gives a direct interpretation of how often a global-crossing improvement agrees with a local-crossing improvement.
\Cref{tab:gcn_lcn_rank_robustness} additionally reports Kendall's $\tau$, robustness to using medians instead of means for the aggregate rankings, and the corresponding aggregate-correlation $p$-values.

\begin{table}[p]
    \begin{minipage}[b]{\textwidth}
        \centering
        \caption{Aggregate rankings of non-RL algorithms by mean global and local crossing number. Rank 1 is best, i.e., lowest mean crossing number. These are the rankings used for the aggregate rank correlations.}
        \label{tab:aggregate_gcn_lcn}
        \begin{tabular}{llrrrr}
            \toprule
            Dataset & Algorithm & Mean global cr. & Global rank & Mean local cr. & Local rank \\
            \midrule
            Rome & EI        & 16.950  & 1 & 3.505  & 2 \\
            Rome & VM        & 18.522  & 2 & 2.965  & 1 \\
            Rome & SmartGD   & 22.494  & 3 & 3.680  & 3 \\
            Rome & $(\mathrm{SGD})^2$ & 25.135  & 4 & 3.996  & 4 \\
            Rome & TB        & 30.106  & 5 & 4.047  & 5 \\
            Rome & KK        & 36.102  & 6 & 4.082  & 6 \\
            Rome & FR        & 45.066  & 7 & 5.096  & 7 \\
            \midrule
            BA & EI          & 22.516   & 1 & 3.681  & 1 \\
            BA & VM          & 488.056  & 2 & 19.077 & 2 \\
            BA & $(\mathrm{SGD})^2$ & 1299.696 & 3 & 36.458 & 3 \\
            BA & TB          & 1545.716 & 4 & 38.916 & 5 \\
            BA & KK          & 1817.134 & 5 & 38.668 & 4 \\
            BA & SmartGD     & 1923.690 & 6 & 49.314 & 7 \\
            BA & FR          & 1977.542 & 7 & 43.002 & 6 \\
            \midrule
            Rome+BA & EI     & 17.241  & 1 & 3.514  & 1 \\
            Rome+BA & VM     & 105.893 & 2 & 5.963  & 2 \\
            Rome+BA & $(\mathrm{SGD})^2$ & 321.683 & 3 & 11.549 & 3 \\
            Rome+BA & TB     & 382.737 & 4 & 12.160 & 5 \\
            Rome+BA & KK     & 450.489 & 5 & 12.129 & 4 \\
            Rome+BA & SmartGD & 464.838 & 6 & 14.298 & 7 \\
            Rome+BA & FR     & 494.688 & 7 & 13.915 & 6 \\
            \bottomrule
        \end{tabular}
    \end{minipage}
    
    \bigskip
    
    \begin{minipage}[b]{\textwidth}
        \centering
        \caption{Robustness of global/local crossing-number rank agreement for the 7 non-RL algorithms.
            Aggregate correlations compare algorithm rankings obtained from mean or median global/local crossing-number values.
            Graph-wise correlations are computed separately on each graph and then summarized by mean or median over graphs.}
        \label{tab:gcn_lcn_rank_robustness}
        \begin{tabular}{llrrrr}
            \toprule
            Dataset & Setting & Spearman $\rho$ & Kendall $\tau$ & Spearman $p$ & Kendall $p$ \\
            \midrule
            Rome    & Aggregate, mean   & 0.964 & 0.905 & $<0.001$ & 0.003 \\
            Rome    & Aggregate, median & 0.926 & 0.845 & 0.003    & 0.014 \\
            Rome    & Graph-wise, mean  & 0.609 & 0.530 & --       & -- \\
            Rome    & Graph-wise, median& 0.671 & 0.582 & --       & -- \\
            \midrule
            BA      & Aggregate, mean   & 0.929 & 0.810 & 0.003    & 0.011 \\
            BA      & Aggregate, median & 0.821 & 0.714 & 0.023    & 0.030 \\
            BA      & Graph-wise, mean  & 0.635 & 0.545 & --       & -- \\
            BA      & Graph-wise, median& 0.703 & 0.600 & --       & -- \\
            \midrule
            Rome+BA & Aggregate, mean   & 0.929 & 0.810 & 0.003    & 0.011 \\
            Rome+BA & Aggregate, median & 0.896 & 0.816 & 0.006    & 0.017 \\
            Rome+BA & Graph-wise, mean  & 0.615 & 0.533 & --       & -- \\
            Rome+BA & Graph-wise, median& 0.677 & 0.582 & --       & -- \\
            \bottomrule
        \end{tabular}
    \end{minipage}
\end{table}

The aggregate rankings agree strongly, with Spearman's $\rho=0.964$ on \rome and $\rho=0.929$ on \BA; on \rome only \textsf{EI} and \textsf{VM} swap positions, while on \BA the three best algorithms are identical under both criteria.
The relation remains positive but weaker on individual instances, with median graph-wise Spearman correlations of $0.671$ on \rome and $0.703$ on \BA.
Pairwise concordance gives a direct interpretation: in $81.2\%$ of strict comparisons on \rome and $79.6\%$ on \BA, the algorithm with lower global crossing number also has lower local crossing number.
Thus, global crossing minimization is a strong indicator for local crossing-number quality, but the $19.1\%$ discordance rate on the combined data shows that it is not a substitute objective.

\section{Discussion}
\label{sec:discussion}

\subparagraph*{\hyperref[rq:1]{\bfseries \textsf{\textcolor{blue!60!black}{\emph{RQ1}}}}.}
In terms of running time, \textsf{KK}, \textsf{TB}, \textsf{FR}, and \textsf{SmartGD} are fastest; \textsf{RL(LC)} remains moderate, with median running times of 13.09s on \rome and 25.04s on \BA, while \textsf{RL(GC)} is slower, with medians of 21.08s and 45.07s, respectively.
\textsf{(SGD)\textsuperscript{2}} is slower than both RL variants on both benchmark sets, and \textsf{EI} and \textsf{VM} time out on many \BA instances.
Regarding the achieved global and local crossing numbers,
our approaches behave quite differently compared to the other algorithms.
For the global crossing number, \textsf{RL(GC)} is not the best-performing method,
but it is competitive with the middle group of established heuristics:
on the \rome graphs it ranks third, ahead of \textsf{SmartGD} and \textsf{(SGD)\textsuperscript{2}},
while on the \BA graphs it is close to \textsf{TB} among the algorithms without timeouts.
For the local crossing number, \textsf{RL(LC)} is one of the strongest methods in our experiments.

For the following more precise comparison of the global crossing number, we consider only \textsf{RL(GC)},
and, for the local crossing number, we consider only \textsf{RL(LC)}.
Compared with the slower heuristics, \textsf{RL(GC)} remains weaker than \textsf{EI} and \textsf{VM} when these algorithms finish,
but on the \rome graphs it improves over \textsf{SmartGD} and \textsf{(SGD)\textsuperscript{2}} in crossing number.
It is faster than \textsf{(SGD)\textsuperscript{2}} on both benchmark sets and avoids the \textsf{EI}/\textsf{VM} timeouts on \BA.
The pairwise plots support this picture: \textsf{RL(GC)} often beats the middle group of heuristics,
but not the strongest global crossing minimization algorithms.
Thus, \textsf{RL(GC)} is not a replacement for the best global crossing minimization heuristics,
but the new results show that \textsf{RL(GC)} formulation can also produce competitive results for global crossing reduction.

The picture is different for the local crossing number.
On the \rome graphs, \textsf{RL(LC)} has the lowest mean and median values.
It is better than \textsf{VM} on average, has a comparable median running time to \textsf{VM}, and is faster than \textsf{EI} and \textsf{(SGD)\textsuperscript{2}}.
On the \BA graphs, \textsf{EI} achieves the best local-crossing values on the subset it solves, and \textsf{VM} is also among the strongest methods on its solved instances,
but both suffer from timeouts on the full test set.
Among the algorithms without timeouts on all \BA instances, \textsf{RL(LC)} has the best mean and median local crossing number,
ahead of \textsf{RL(GC)}, \textsf{(SGD)\textsuperscript{2}}, \textsf{KK}, \textsf{TB}, \textsf{FR}, and \textsf{SmartGD}.
The pairwise plots reinforce this distinction between raw solution quality on solved instances and robustness on the full \BA test set, where timeouts count against the timed-out algorithm.
It is also a little more than twice as fast as \textsf{(SGD)\textsuperscript{2}} in median running time on the \BA graphs.

We conclude that our RL approach is a strong choice for optimizing the local crossing number,
in particular for the tested graph sizes (up to about 150 vertices).
For the global crossing number it is not the best method, but it is competitive with several established heuristics.
This recommendation should be taken with caution since
the benchmark competitors do not explicitly optimize the local crossing number.

\subparagraph*{\hyperref[rq:2]{\bfseries \textsf{\textcolor{blue!60!black}{\emph{RQ2}}}}.}
Comparing the performance of the existing heuristics ignoring our RL methods,
our experiments indicate a strong correlation between performance with respect to global crossing number and performance with respect to local crossing number; see \cref{tab:gcn_lcn_agreement}.
This holds both for aggregate rankings over the benchmark sets and, more moderately, for graph-wise rankings.
Thus, the evaluated non-RL algorithms that perform well for global crossing minimization are usually also good choices for obtaining low local crossing number.
However, the correlation is not perfect, and global crossing number is not a substitute objective for local crossing number.
The clearest example is the pair \textsf{EI}/\textsf{VM}: on the \rome graphs, \textsf{EI} has the better mean global crossing number, whereas \textsf{VM} has the better mean local crossing number; see \cref{tab:aggregate_gcn_lcn}.

The discordant pairwise comparisons reported in \cref{tab:gcn_lcn_agreement} further show that explicitly optimizing the local objective can matter on individual instances.
The performance of \textsf{RL(LC)} supports the same conclusion: optimizing the local objective directly can improve over methods that only optimize global crossing number.

\subparagraph*{Limitations.}
We trained our models only on two graph classes and cannot infer from the results that our approach generalizes to arbitrary graphs.
Additional training and experimentation on different graph classes might be necessary to establish generalizability.
While the running times of our approaches are better than the slowest competitors and avoid the timeouts observed for \textsf{EI} and \textsf{VM} on the \BA graphs,
they are still slower than the fastest heuristics, resulting in limited scalability for larger graphs.
We believe this might be remedied by a more efficient implementation, e.g., implemented entirely in C++.
Finally, in our experiments, we restricted ourselves to a single initial layout method, and it remains unclear how strongly performance depends on it, i.e.,
can we expect similar performance
on structurally very different initial layouts?
Reducing the local crossing number on the grid is precisely the task in the annual Graph Drawing Contest in 2025 and 2026, but we did not compare against last year's challenge algorithms because their code is not publicly available.

\section{Conclusion and Future Work}
\label{sec:conclusion}
We have formulated the problem of generating straight-line graph drawings as a game,
for which we propose an RL approach.
This can be seen as a novel post-processing procedure
to optimize the global and the local crossing number
of a given initial layout generated by a state-of-the-art graph-drawing algorithm.
We explicitly optimize the local crossing number,
for which \textsf{RL(LC)} is one of the strongest methods in our experiments:
it performs best on the \rome graphs and best among the algorithms that solve all \BA instances within the time limit.
For the global crossing number, \textsf{RL(GC)} does not match the best specialized heuristics,
but it ranks third on the \rome graphs, remains competitive with several established methods on \BA, and is substantially faster than some slower competitors.
We conclude that repair-based reinforcement learning
is a viable way to optimize the local crossing number directly
where existing global-crossing heuristics are useful but not always sufficient.
Moreover, it might be a promising framework for optimizing graph layouts
in general,
but the modeling of the repair state, vertex selection, and reward remains crucial for each target objective.

Future work includes training on more graph classes, improving scalability, starting with other initial layout algorithms, and exploring alternative observation spaces, action spaces, and RL frameworks.
We also plan to test contest instances once this becomes feasible; at present this is limited by
\begin{inparaenum}[(1)]
    \item unavailable benchmark instances and implementations,
    \item our agent is not yet tuned to work on small grids,
    \item and our training needs to be extended to graphs that are structurally different from Rome and Barabási-Albert graphs.
\end{inparaenum}
Finally, the same repair-based idea could be tested for other quantifiable drawing objectives.

\bibliographystyle{plainurl}
\bibliography{bibliography}

\begin{thebibliography}{10}

\bibitem{ahmed2022multicriteria}
Reyan Ahmed, Felice De~Luca, Sabin Devkota, Stephen Kobourov, and Mingwei Li.
\newblock Multicriteria scalable graph drawing via stochastic gradient descent,
  {(SGD)}$^2$.
\newblock {\em IEEE Transactions on Visualization and Computer Graphics},
  28(6):2388--2399, 2022.
\newblock \href {https://doi.org/10.1109/TVCG.2022.3155564}
  {\path{doi:10.1109/TVCG.2022.3155564}}.

\bibitem{albert2000topology}
R{\'e}ka Albert and Albert-L{\'a}szl{\'o} Barab{\'a}si.
\newblock Topology of evolving networks: local events and universality.
\newblock {\em Physical review letters}, 85(24):5234, 2000.
\newblock \href {https://doi.org/10.1103/physrevlett.85.5234}
  {\path{doi:10.1103/physrevlett.85.5234}}.

\bibitem{DBLP:journals/cj/ArgyriouBS13}
Evmorfia~N. Argyriou, Michael~A. Bekos, and Antonios Symvonis.
\newblock Maximizing the total resolution of graphs.
\newblock {\em Comput. J.}, 56(7):887--900, 2013.
\newblock \href {https://doi.org/10.1093/COMJNL/BXS088}
  {\path{doi:10.1093/COMJNL/BXS088}}.

\bibitem{DBLP:journals/cgf/BartolomeoCSPWD24}
Sara~Di Bartolomeo, Tarik Crnovrsanin, David Saffo, Eduardo Puerta, Connor
  Wilson, and Cody Dunne.
\newblock Evaluating graph layout algorithms: {A} systematic review of methods
  and best practices.
\newblock {\em Comput. Graph. Forum}, 43(6), 2024.
\newblock \href {https://doi.org/10.1111/CGF.15073}
  {\path{doi:10.1111/CGF.15073}}.

\bibitem{DBLP:journals/jss/BatiniTT84}
Carlo Batini, Maurizio Talamo, and Roberto Tamassia.
\newblock Computer aided layout of entity relationship diagrams.
\newblock {\em J. Syst. Softw.}, 4(2--3):163--173, 1984.
\newblock \href {https://doi.org/10.1016/0164-1212(84)90006-2}
  {\path{doi:10.1016/0164-1212(84)90006-2}}.

\bibitem{DBLP:journals/cj/BekosFGHKSS21}
Michael~A. Bekos, Henry F{\"{o}}rster, Christian Geckeler, Lukas
  Holl{\"{a}}nder, Michael Kaufmann, Amad{\"{a}}us~M. Spallek, and Jan Splett.
\newblock A heuristic approach towards drawings of graphs with high crossing
  resolution.
\newblock {\em Comput. J.}, 64(1):7--26, 2021.
\newblock \href {https://doi.org/10.1093/COMJNL/BXZ133}
  {\path{doi:10.1093/COMJNL/BXZ133}}.

\bibitem{bienstock}
Daniel Bienstock.
\newblock Some provably hard crossing number problems.
\newblock {\em Discrete \& Computational Geometry}, 6(3):443--459, 1991.
\newblock \href {https://doi.org/10.1007/BF02574701}
  {\path{doi:10.1007/BF02574701}}.

\bibitem{DBLP:conf/gd/ChimaniDR18}
Markus Chimani, Hanna D{\"{o}}ring, and Matthias Reitzner.
\newblock Crossing numbers and stress of random graphs.
\newblock In Therese Biedl and Andreas Kerren, editors, {\em Proc. 26th
  International Symposium on Graph Drawing and Network Visualization ({GD}
  2018)}, volume 11282 of {\em Lecture Notes in Computer Science}, pages
  255--268. Springer, 2018.
\newblock \href {https://doi.org/10.1007/978-3-030-04414-5\_18}
  {\path{doi:10.1007/978-3-030-04414-5\_18}}.

\bibitem{DBLP:conf/esa/ChimaniW16}
Markus Chimani and Tilo Wiedera.
\newblock An {ILP}-based proof system for the crossing number problem.
\newblock In Piotr Sankowski and Christos~D. Zaroliagis, editors, {\em Proc.
  24th Annual European Symposium on Algorithms (ESA2016)}, volume~57 of {\em
  LIPIcs}, pages 29:1--29:13. Schloss Dagstuhl - Leibniz-Zentrum f{\"{u}}r
  Informatik, 2016.
\newblock \href {https://doi.org/10.4230/LIPICS.ESA.2016.29}
  {\path{doi:10.4230/LIPICS.ESA.2016.29}}.

\bibitem{amzplayer2023}
P.~David.
\newblock {AmzPlayer}: A modern player for {Z}illions.
\newblock \url{http://www.polyomino.com/amzplayer/}, 2023.

\bibitem{DBLP:conf/gd/DemelDMRW18}
Almut Demel, Dominik D{\"{u}}rrschnabel, Tamara Mchedlidze, Marcel Radermacher,
  and Lasse Wulf.
\newblock A greedy heuristic for crossing-angle maximization.
\newblock In Therese Biedl and Andreas Kerren, editors, {\em Proc. 26th
  International Symposium on Graph Drawing and Network Visualization (GD
  2018)}, volume 11282 of {\em Lecture Notes in Computer Science}, pages
  286--299. Springer, 2018.
\newblock \href {https://doi.org/10.1007/978-3-030-04414-5\_20}
  {\path{doi:10.1007/978-3-030-04414-5\_20}}.

\bibitem{DBLP:conf/gd/DevkotaALIK19}
Sabin Devkota, Abu~Reyan Ahmed, Felice~De Luca, Katherine~E. Isaacs, and
  Stephen~G. Kobourov.
\newblock Stress-plus-{X} {(SPX)} graph layout.
\newblock In Daniel Archambault and Csaba~D. T{\'{o}}th, editors, {\em Proc.
  27th International Symposium on Graph Drawing and Network Visualization (GD
  2019)}, volume 11904 of {\em Lecture Notes in Computer Science}, pages
  291--304. Springer, 2019.
\newblock \href {https://doi.org/10.1007/978-3-030-35802-0\_23}
  {\path{doi:10.1007/978-3-030-35802-0\_23}}.

\bibitem{di1997experimental}
Giuseppe Di~Battista, Ashim Garg, Giuseppe Liotta, Roberto Tamassia, Emanuele
  Tassinari, and Francesco Vargiu.
\newblock An experimental comparison of four graph drawing algorithms.
\newblock {\em Computational Geometry}, 7(5--6):303--325, 1997.
\newblock \href {https://doi.org/10.1016/s0925-7721(96)00005-3}
  {\path{doi:10.1016/s0925-7721(96)00005-3}}.

\bibitem{DBLP:journals/csur/DidimoLM19}
Walter Didimo, Giuseppe Liotta, and Fabrizio Montecchiani.
\newblock A survey on graph drawing beyond planarity.
\newblock {\em {ACM} Comput. Surv.}, 52(1):4:1--4:37, 2019.
\newblock \href {https://doi.org/10.1145/3301281} {\path{doi:10.1145/3301281}}.

\bibitem{DBLP:conf/gd/DidimoLR10}
Walter Didimo, Giuseppe Liotta, and Salvatore~Agostino Romeo.
\newblock Topology-driven force-directed algorithms.
\newblock In Ulrik Brandes and Sabine Cornelsen, editors, {\em Proc. 18th
  International Symposium on Graph Drawing (GD 2010)}, volume 6502 of {\em
  Lecture Notes in Computer Science}, pages 165--176. Springer, 2010.
\newblock \href {https://doi.org/10.1007/978-3-642-18469-7\_15}
  {\path{doi:10.1007/978-3-642-18469-7\_15}}.

\bibitem{muzero-general}
Werner Duvaud and Aurèle Hainaut.
\newblock Muzero general: Open reimplementation of muzero.
\newblock \url{https://github.com/werner-duvaud/muzero-general}, 2019.

\bibitem{fruchterman1991graph}
Thomas~M.J. Fruchterman and Edward~M. Reingold.
\newblock Graph drawing by force-directed placement.
\newblock {\em Software: Practice and experience}, 21(11):1129--1164, 1991.
\newblock \href {https://doi.org/10.1002/spe.4380211102}
  {\path{doi:10.1002/spe.4380211102}}.

\bibitem{doi:10.1137/0604033}
M.~R. Garey and D.~S. Johnson.
\newblock Crossing number is {NP}-complete.
\newblock {\em SIAM Journal on Algebraic Discrete Methods}, 4(3):312--316,
  1983.
\newblock \href {https://doi.org/10.1137/0604033} {\path{doi:10.1137/0604033}}.

\bibitem{DBLP:conf/gd/GiovannangeliLA21}
Loann Giovannangeli, Fr{\'{e}}d{\'{e}}ric Lalanne, David Auber, Romain Giot,
  and Romain Bourqui.
\newblock Deep neural network for drawing networks, {$(DNN)^2$}.
\newblock In Helen~C. Purchase and Ignaz Rutter, editors, {\em Proc. 29th
  International Symposium on Graph Drawing and Network Visualization (GD
  2021)}, volume 12868 of {\em Lecture Notes in Computer Science}, pages
  375--390. Springer, 2021.
\newblock \href {https://doi.org/10.1007/978-3-030-92931-2\_27}
  {\path{doi:10.1007/978-3-030-92931-2\_27}}.

\bibitem{DBLP:journals/algorithmica/GrigorievB07}
Alexander Grigoriev and Hans~L. Bodlaender.
\newblock Algorithms for graphs embeddable with few crossings per edge.
\newblock {\em Algorithmica}, 49(1):1--11, 2007.
\newblock \href {https://doi.org/10.1007/S00453-007-0010-X}
  {\path{doi:10.1007/S00453-007-0010-X}}.

\bibitem{hagberg2008exploring}
Aric Hagberg, Pieter~J Swart, and Daniel~A Schult.
\newblock Exploring network structure, dynamics, and function using networkx.
\newblock Technical report, Los Alamos National Laboratory (LANL), Los Alamos,
  NM (United States), 2008.
\newblock Function used to generate extended Barabási-Albert graphs:
  \url{https://networkx.org/documentation/stable/reference/generated/networkx.generators.random_graphs.extended_barabasi_albert_graph.html}.

\bibitem{HuangEHL13}
Weidong Huang, Peter Eades, Seok{-}Hee Hong, and Chun{-}Cheng Lin.
\newblock Improving multiple aesthetics produces better graph drawings.
\newblock {\em J. Vis. Lang. Comput.}, 24(4):262--272, 2013.
\newblock \href {https://doi.org/10.1016/J.JVLC.2011.12.002}
  {\path{doi:10.1016/J.JVLC.2011.12.002}}.

\bibitem{kamada1989algorithm}
Tomihisa Kamada and Satoru Kawai.
\newblock An algorithm for drawing general undirected graphs.
\newblock {\em Information Processing Letters}, 31(1):7--15, 1989.
\newblock \href {https://doi.org/10.1016/0020-0190(89)90102-6}
  {\path{doi:10.1016/0020-0190(89)90102-6}}.

\bibitem{DBLP:journals/csr/KobourovLM17}
Stephen~G. Kobourov, Giuseppe Liotta, and Fabrizio Montecchiani.
\newblock An annotated bibliography on 1-planarity.
\newblock {\em Comput. Sci. Rev.}, 25:49--67, 2017.
\newblock \href {https://doi.org/10.1016/J.COSREV.2017.06.002}
  {\path{doi:10.1016/J.COSREV.2017.06.002}}.

\bibitem{DBLP:conf/gd/KorzhikM08}
Vladimir~P. Korzhik and Bojan Mohar.
\newblock Minimal obstructions for 1-immersions and hardness of 1-planarity
  testing.
\newblock In Ioannis~G. Tollis and Maurizio Patrignani, editors, {\em Proc.
  16th International Symposium on Graph Drawing (GD 2008)}, volume 5417 of {\em
  Lecture Notes in Computer Science}, pages 302--312. Springer, 2008.
\newblock \href {https://doi.org/10.1007/978-3-642-00219-9\_29}
  {\path{doi:10.1007/978-3-642-00219-9\_29}}.

\bibitem{DBLP:journals/jgt/KorzhikM13}
Vladimir~P. Korzhik and Bojan Mohar.
\newblock Minimal obstructions for 1-immersions and hardness of 1-planarity
  testing.
\newblock {\em J. Graph Theory}, 72(1):30--71, 2013.
\newblock \href {https://doi.org/10.1002/JGT.21630}
  {\path{doi:10.1002/JGT.21630}}.

\bibitem{DBLP:journals/tvcg/KwonM20}
Oh{-}Hyun Kwon and Kwan{-}Liu Ma.
\newblock A deep generative model for graph layout.
\newblock {\em {IEEE} Trans. Vis. Comput. Graph.}, 26(1):665--675, 2020.
\newblock \href {https://doi.org/10.1109/TVCG.2019.2934396}
  {\path{doi:10.1109/TVCG.2019.2934396}}.

\bibitem{LanctotEtAl2019OpenSpiel}
Marc Lanctot, Edward Lockhart, Jean-Baptiste Lespiau, Vinicius Zambaldi,
  Satyaki Upadhyay, Julien P\'{e}rolat, Sriram Srinivasan, Finbarr Timbers,
  Karl Tuyls, Shayegan Omidshafiei, Daniel Hennes, Dustin Morrill, Paul Muller,
  Timo Ewalds, Ryan Faulkner, J\'{a}nos Kram\'{a}r, Bart~De Vylder, Brennan
  Saeta, James Bradbury, David Ding, Sebastian Borgeaud, Matthew Lai, Julian
  Schrittwieser, Thomas Anthony, Edward Hughes, Ivo Danihelka, and Jonah
  Ryan-Davis.
\newblock {OpenSpiel}: {A} framework for reinforcement learning in games.
\newblock {\em ArXiv preprint, CoRR}, 2019.
\newblock \href {https://arxiv.org/abs/1908.09453} {\path{arXiv:1908.09453}}.

\bibitem{DBLP:journals/nature/MnihKSRVBGRFOPB15}
Volodymyr Mnih, Koray Kavukcuoglu, David Silver, Andrei~A. Rusu, Joel Veness,
  Marc~G. Bellemare, Alex Graves, Martin~A. Riedmiller, Andreas Fidjeland,
  Georg Ostrovski, Stig Petersen, Charles Beattie, Amir Sadik, Ioannis
  Antonoglou, Helen King, Dharshan Kumaran, Daan Wierstra, Shane Legg, and
  Demis Hassabis.
\newblock Human-level control through deep reinforcement learning.
\newblock {\em Nature}, 518(7540):529--533, 2015.
\newblock \href {https://doi.org/10.1038/NATURE14236}
  {\path{doi:10.1038/NATURE14236}}.

\bibitem{DBLP:journals/corr/abs-1912-06680}
OpenAI, Christopher Berner, Greg Brockman, Brooke Chan, Vicki Cheung,
  Przemysław Dębiak, Christy Dennison, David Farhi, Quirin Fischer, Shariq
  Hashme, Chris Hesse, Rafal Józefowicz, Scott Gray, Catherine Olsson, Jakub
  Pachocki, Michael Petrov, Henrique~Pondé de~Oliveira~Pinto, Jonathan Raiman,
  Tim Salimans, Jeremy Schlatter, Jonas Schneider, Szymon Sidor, Ilya
  Sutskever, Jie Tang, Filip Wolski, and Susan Zhang.
\newblock Dota 2 with large scale deep reinforcement learning.
\newblock {\em ArXiv preprint, CoRR}, 2019.
\newblock \href {https://arxiv.org/abs/1912.06680} {\path{arXiv:1912.06680}}.

\bibitem{maxi}
Maximilian Pfister.
\newblock Personal communication, 2025.

\bibitem{Purchase2000}
Helen~C. Purchase.
\newblock Effective information visualisation: a study of graph drawing
  aesthetics and algorithms.
\newblock {\em Interacting with Computers}, 13(2):147--162, 2000.
\newblock \href {https://doi.org/10.1016/S0953-5438(00)00032-1}
  {\path{doi:10.1016/S0953-5438(00)00032-1}}.

\bibitem{Purchase2002}
Helen~C. Purchase, David~A. Carrington, and Jo{-}Anne Allder.
\newblock Empirical evaluation of aesthetics-based graph layout.
\newblock {\em Empirical Software Engineering}, 7(3):233--255, 2002.
\newblock \href {https://doi.org/10.1023/A:1016344215610}
  {\path{doi:10.1023/A:1016344215610}}.

\bibitem{radermacher2019geometric}
Marcel Radermacher, Klara Reichard, Ignaz Rutter, and Dorothea Wagner.
\newblock Geometric heuristics for rectilinear crossing minimization.
\newblock {\em {ACM} J. Exp. Algorithmics}, 24(1):1.12:1--1.12:21, 2019.
\newblock \href {https://doi.org/10.1145/3325861} {\path{doi:10.1145/3325861}}.

\bibitem{DBLP:journals/corr/abs-2011-00748}
Ilkin Safarli, Youjia Zhou, and Bei Wang.
\newblock Interpreting graph drawing with multi-agent reinforcement learning.
\newblock {\em ArXiv preprint, CoRR}, 2020.
\newblock \href {https://arxiv.org/abs/2011.00748} {\path{arXiv:2011.00748}}.

\bibitem{survey-crossing-number}
Marcus Schaefer.
\newblock The graph crossing number and its variants: A survey.
\newblock {\em The Electronic Journal of Combinatorics}, 2013 (last updated in
  2024).
\newblock \href {https://doi.org/10.37236/2713} {\path{doi:10.37236/2713}}.

\bibitem{DBLP:journals/jgaa/Schaefer21}
Marcus Schaefer.
\newblock Complexity of geometric $k$-planarity for fixed $k$.
\newblock {\em J. Graph Algorithms Appl.}, 25(1):29--41, 2021.
\newblock \href {https://doi.org/10.7155/JGAA.00548}
  {\path{doi:10.7155/JGAA.00548}}.

\bibitem{DBLP:journals/jgaa/Schaefer21a}
Marcus Schaefer.
\newblock On the complexity of some geometric problems with fixed parameters.
\newblock {\em J. Graph Algorithms Appl.}, 25(1):195--218, 2021.
\newblock \href {https://doi.org/10.7155/JGAA.00557}
  {\path{doi:10.7155/JGAA.00557}}.

\bibitem{DBLP:journals/nature/SchrittwieserAH20}
Julian Schrittwieser, Ioannis Antonoglou, Thomas Hubert, Karen Simonyan,
  Laurent Sifre, Simon Schmitt, Arthur Guez, Edward Lockhart, Demis Hassabis,
  Thore Graepel, Timothy~P. Lillicrap, and David Silver.
\newblock Mastering {Atari}, {Go}, chess and shogi by planning with a learned
  model.
\newblock {\em Nature}, 588(7839):604--609, 2020.
\newblock \href {https://doi.org/10.1038/S41586-020-03051-4}
  {\path{doi:10.1038/S41586-020-03051-4}}.

\bibitem{DBLP:journals/corr/SchulmanWDRK17}
John Schulman, Filip Wolski, Prafulla Dhariwal, Alec Radford, and Oleg Klimov.
\newblock Proximal policy optimization algorithms.
\newblock {\em ArXiv preprint, CoRR}, 2017.
\newblock \href {https://arxiv.org/abs/1707.06347} {\path{arXiv:1707.06347}}.

\bibitem{DBLP:journals/nature/SilverHMGSDSAPL16}
David Silver, Aja Huang, Chris~J. Maddison, Arthur Guez, Laurent Sifre, George
  van~den Driessche, Julian Schrittwieser, Ioannis Antonoglou, Vedavyas
  Panneershelvam, Marc Lanctot, Sander Dieleman, Dominik Grewe, John Nham, Nal
  Kalchbrenner, Ilya Sutskever, Timothy~P. Lillicrap, Madeleine Leach, Koray
  Kavukcuoglu, Thore Graepel, and Demis Hassabis.
\newblock Mastering the game of {Go} with deep neural networks and tree search.
\newblock {\em Nature}, 529(7587):484--489, 2016.
\newblock \href {https://doi.org/10.1038/NATURE16961}
  {\path{doi:10.1038/NATURE16961}}.

\bibitem{doi:10.1126/science.aar6404}
David Silver, Thomas Hubert, Julian Schrittwieser, Ioannis Antonoglou, Matthew
  Lai, Arthur Guez, Marc Lanctot, Laurent Sifre, Dharshan Kumaran, Thore
  Graepel, Timothy Lillicrap, Karen Simonyan, and Demis Hassabis.
\newblock A general reinforcement learning algorithm that masters chess, shogi,
  and {Go} through self-play.
\newblock {\em Science}, 362(6419):1140--1144, 2018.
\newblock \href {https://doi.org/10.1126/science.aar6404}
  {\path{doi:10.1126/science.aar6404}}.

\bibitem{DBLP:journals/nature/SilverSSAHGHBLB17}
David Silver, Julian Schrittwieser, Karen Simonyan, Ioannis Antonoglou, Aja
  Huang, Arthur Guez, Thomas Hubert, Lucas Baker, Matthew Lai, Adrian Bolton,
  Yutian Chen, Timothy~P. Lillicrap, Fan Hui, Laurent Sifre, George van~den
  Driessche, Thore Graepel, and Demis Hassabis.
\newblock Mastering the game of {Go} without human knowledge.
\newblock {\em Nature}, 550(7676):354--359, 2017.
\newblock \href {https://doi.org/10.1038/NATURE24270}
  {\path{doi:10.1038/NATURE24270}}.

\bibitem{Tavener2025}
Stephen Tavener.
\newblock {Ai Ai}: a {J}ava-based general game playing engine.
\newblock \url{http://mrraow.com/index.php/aiai-home/}, 2025.

\bibitem{crossing-numbers}
{Theoretical Computer Science Group, University of Osnabr\"uck}.
\newblock Crossing number~-- {R}ome graphs: Exact crossing numbers.
\newblock
  \url{https://tcs.uos.de/doku.php?id=research/cr#rome_graphsexact_crossing_numbers},
  2025.

\bibitem{DBLP:journals/cga/WangYHS21}
Xiaoqi Wang, Kevin Yen, Yifan Hu, and Han{-}Wei Shen.
\newblock {DeepGD:} {A} deep learning framework for graph drawing using {GNN}.
\newblock {\em {IEEE} Computer Graphics and Applications}, 41(5):32--44, 2021.
\newblock \href {https://doi.org/10.1109/MCG.2021.3093908}
  {\path{doi:10.1109/MCG.2021.3093908}}.

\bibitem{wang2023smartgd}
Xiaoqi Wang, Kevin Yen, Yifan Hu, and Han-Wei Shen.
\newblock {SmartGD}: A {GAN}-based graph drawing framework for diverse
  aesthetic goals.
\newblock {\em IEEE Transactions on Visualization and Computer Graphics},
  30(8):5666--5678, 2023.
\newblock \href {https://doi.org/10.1109/tvcg.2023.3306356}
  {\path{doi:10.1109/tvcg.2023.3306356}}.

\bibitem{DBLP:journals/tvcg/WangJWCMQ20}
Yong Wang, Zhihua Jin, Qianwen Wang, Weiwei Cui, Tengfei Ma, and Huamin Qu.
\newblock Deepdrawing: {A} deep learning approach to graph drawing.
\newblock {\em {IEEE} Trans. Vis. Comput. Graph.}, 26(1):676--686, 2020.
\newblock \href {https://doi.org/10.1109/TVCG.2019.2934798}
  {\path{doi:10.1109/TVCG.2019.2934798}}.

\bibitem{Ware2002}
Colin Ware, Helen~C. Purchase, Linda Colpoys, and Matthew McGill.
\newblock Cognitive measurements of graph aesthetics.
\newblock {\em Information Visualization}, 1(2):103--110, 2002.
\newblock \href {https://doi.org/10.1057/palgrave.ivs.9500013}
  {\path{doi:10.1057/palgrave.ivs.9500013}}.

\bibitem{WatkinsD92}
Christopher J. C.~H. Watkins and Peter Dayan.
\newblock {Q}-learning (technical note).
\newblock {\em Mach. Learn.}, 8:279--292, 1992.
\newblock \href {https://doi.org/10.1007/BF00992698}
  {\path{doi:10.1007/BF00992698}}.

\end{thebibliography}

\end{document}